\documentclass[preprint]{aastex}
\shorttitle{A Study of A262}
\shortauthors{Neill et al.}

\def\about{ \ifmmode{\sim} \else $\sim$ \fi}
\def\prop{ \ifmmode{\propto} \else $\propto$ \fi}
\def\sigp{ \ifmmode{\sigma_p} \else $\sigma_p$ \fi}
\def\sigr{ \ifmmode{\sigma_r} \else $\sigma_r$ \fi}
\def\sigth{ \ifmmode{\sigma_\theta} \else $\sigma_\theta$ \fi}
\def\bspec{ \ifmmode{\beta_{spec}} \else $\beta_{spec}$ \fi}
\def\bgas{ \ifmmode{\beta_{fit,gas}} \else $\beta_{fit,gas}$ \fi}
\def\bgal{ \ifmmode{\beta_{fit,gal}} \else $\beta_{fit,gal}$ \fi}
\def\bfit{ \ifmmode{\beta_{fit}} \else $\beta_{fit}$ \fi}
\def\bfitc{ \ifmmode{\beta_{fit}^c} \else $\beta_{fit}^c$ \fi}
\def\tx{ \ifmmode{T_x} \else $T_x$ \fi}
\def\rhogas{ \ifmmode{\rho_{gas}} \else $\rho_{gas}$ \fi}
\def\rhogal{ \ifmmode{\rho_{gal}} \else $\rho_{gal}$ \fi}
\def\rhoc{ \ifmmode{\rho_c} \else $\rho_c$ \fi}

\def\omega0{ \ifmmode{\Omega_0} \else $\Omega_0$ \fi}
\def\H0{ \ifmmode{H_0} \else $H_0$ \fi}
\def\ang{\AA}
\def\kms{km~s$^{-1}$}
\def\degpnt{^{\circ}\kern-1.7mm.\kern+.35mm}
\def\deg{^{\circ}}

\begin{document}

\title{The Beta Problem: A Study of Abell 262}

\author{James D. Neill\altaffilmark{1}}
\affil{Astronomy Department, Columbia University, New York, NY 10027}
\email{neill@astro.columbia.edu}

\author{Jean P. Brodie}
\affil{Lick Observatory, University of California, Santa Cruz, CA 95064}
\email{brodie@ucolick.org}

\author{William W. Craig\altaffilmark{1} and Charles J. Hailey}
\affil{Columbia Astrophysics Laboratory, Columbia University, New York,
	NY, 10027}
\email{bill@astro.columbia.edu}
\email{chuckh@astro.columbia.edu}

\and

\author{Anthony A. Misch}
\affil{Lick Observatory, University of California, Santa Cruz, CA 95064}
\email{tony@ucolick.org}

\altaffiltext{1}{Visiting Astronomer, Lick Observatory.
Lick Observatory is operated by the Regents of the University of
California.}

\begin{abstract}

We present an investigation of the dynamical state of the cluster
A262.  Existing optical line of sight velocities for select cluster
galaxies have been augmented by new data obtained with the Automated
Multi-Object Spectrograph at Lick Observatory.  We find evidence 
for a virialized early-type population distinct from a late-type 
population infalling from the Pisces-Perseus supercluster ridge.
We also report on a tertiary population of low luminosity galaxies
whose velocity dispersion distinguishes them from both the early 
and late-type galaxies.  We supplement our investigation with an
analysis of archival X-ray data.  A temperature is determined using 
{\it ASCA} GIS data and a gas profile is derived from {\it ROSAT} HRI data.

The increased statistics of our sample results in a picture of A262
with significant differences from earlier work.  A previously proposed
solution to the ``$\beta$--problem'' in A262 in which the gas temperature is
significantly higher than the galaxy temperature is shown to result
from using too low a velocity dispersion for the early-type galaxies.
Our data present a consistent picture of A262 in which there
is no ``$\beta$--problem'', and the gas and galaxy temperature are roughly
comparable. There is no longer any requirement for extensive galaxy-gas
feedback to drastically overheat the gas with respect to the galaxies. 
We also demonstrate that entropy-floor models can explain the recent
discovery that the $\beta$ values determined by cluster gas and the  
cluster core radii are correlated.

\end{abstract}


\section{INTRODUCTION}
\label{intro}

Clusters of galaxies provide an opportunity to study the formation of
massive structures through gravitational accretion. The cluster dark matter
potential well is an environment in which infalling and virialized
galaxies interact with intra-cluster gas in a complex, time-dependent
fashion involving dynamical friction with dark matter as well as galactic
winds and ram-pressure stripping.  Measurements of cluster galaxies and gas
temperatures as a function of radius, element abundances,
luminosity-temperature relations and the evolution of these quantities with
redshift all provide data to test theories of cluster formation and
evolution.  Clusters also provide a probe of many cosmological
parameters.  The density of clusters forming as a function of redshift can be
used to measure quantities such as \omega0 and the spectrum of
density perturbations in the early universe.

One problem which arises in exploring the topics described above is
the $\beta$--problem. Assuming that gas and galaxies are both tracers of the
dark matter potential well, and that no source of energy other than gravity
is operative, there is a relation between the ratio of the
internal energy of the galaxies and the gas temperature (determined using
spectroscopic data and generally termed \bspec) to a quantity related to the
gas and galaxy density profiles (determined by fitting isophotes to the X-ray
gas and optical emission and generally termed \bfit).  The failure of these
two quantities to agree in most clusters was noted some time ago \citep{sar86}.
Numerous solutions have been offered over the years.  These include
adjustments of \bfit to properly account for the galaxy distribution
as indicated by galaxy-galaxy correlation measurements
\citep{bah94}; galaxy velocity distribution asymmetries \citep{men86}; 
incorrect determinations of \bspec due to contamination of velocity 
dispersions by cluster substructure \citep{edg91b};
and presence of significant galaxy-gas feedback or
dynamical friction with dark matter \citep{bir95}.  
\citet{dav96} have proposed
that the $\beta$--problem is solved by excluding late-type galaxies
when estimating the cluster velocity dispersion since such galaxies are still
infalling and therefore not virialized.

The $\beta$--problem is significant for several reasons.  The measurement of
$\beta$ from X-ray and optical data provides information about the nature
of galaxy-gas feedback. It is complementary to other methods used to probe
the galaxy-gas interaction such as heavy element abundance gradients.
The $\beta$--problem will become increasingly important as more
accurate cluster temperatures become available from {\it Chandra} and 
{\it XMM}.  With improved temperature measurements, $\beta$ uncertainty, which
appears in the conversion from gas temperature to mass in the
Press-Schechter formula \citep{pre74}, can contribute significantly 
to the mass
uncertainty. It is essential to confirm that the gas dynamic-dark matter
simulations are correct, since they are utilized to obtain $\beta$.  In fact,
the comparison of $\beta$ values derived from modeling with those obtained
from optical observations indicate marginal agreement given the observational
and theoretical uncertainties \citep{hen97}.

Many problems can lead to a cluster violating
the assumptions required for \bspec to equal
\bfit.  These problems include use of non-virialized galaxy populations, use
of galaxies associated with infall or other peculiarities in the velocity
field, local sub-clustering, presence of cooling flows or magnetic fields
etc..  There are several approaches to evaluating the
possibilities.  The first approach is to observe many clusters, with the
number of spectra collected per cluster generally limited to $\about20$ 
or less, and
to statistically analyze the resultant data.  With adequate temperatures and
substructure analyses the calculation of ``average''
cluster properties may then represent a ``typical'' member of the
sample population, with any scatter being random.  As a complement to this
statistical approach, an analysis of individual clusters through use
of extensive X-ray and optical data permits correction for many effects and
thus provides an alternative testbed for comparison with both theory and
the results of the statistical approach. Such detailed
analyses are rare because they require much high quality optical and
X-ray data.  We are obtaining large numbers of redshifts for several
clusters for eventual comparison with {\it XMM} and {\it Chandra} data.

In this paper we describe a detailed investigation of A262.  We
augmented existing galaxy redshift data with a sample of new
redshifts obtained with the Automated Multi-Object Spectrograph at Lick
Observatory. We have also analyzed previously unpublished {\it ROSAT}
and {\it ASCA} data.  The complete data set allows us to address the
problems discussed above.

\section{OBSERVATIONS \& INITIAL ANALYSIS} \label{optical}

We combined new observations of A262 with published redshifts to
produce the sample from which we derived cluster optical 
properties.  Here we describe the new observations, compare our data to 
published data, and discuss the generation of a uniform catalog.
We also present our analysis of the X-ray data to derive the profile
and temperature of the X-ray gas.  We use $\H0=65$ \kms Mpc$^{-1}$
in our analysis.

\subsection{AMOS Observations}

The Automated Multi-Object Spectrograph (AMOS, Craig et al. 1993) consists of a
robotic fiber positioner at the prime focus of the Lick 3m telescope, a fiber
feed, and a floor-mounted spectrograph.  Sixty fibers (39 more are
planned) can be positioned within a $1\deg$ field of view to an accuracy of
0.1 arcsecond.  Constraints on the fiber placement typically limit the number
of fibers that can be assigned to objects and the unassigned fibers are used
to monitor the sky background.  We used the red channel of the spectrograph
with the 700 line/mm grating giving a spectral resolution of 0.7 \ang/pixel
over the range 6200 -- 7500 \ang.

\begin{figure}[h]
\epsscale{0.5}
\plotone{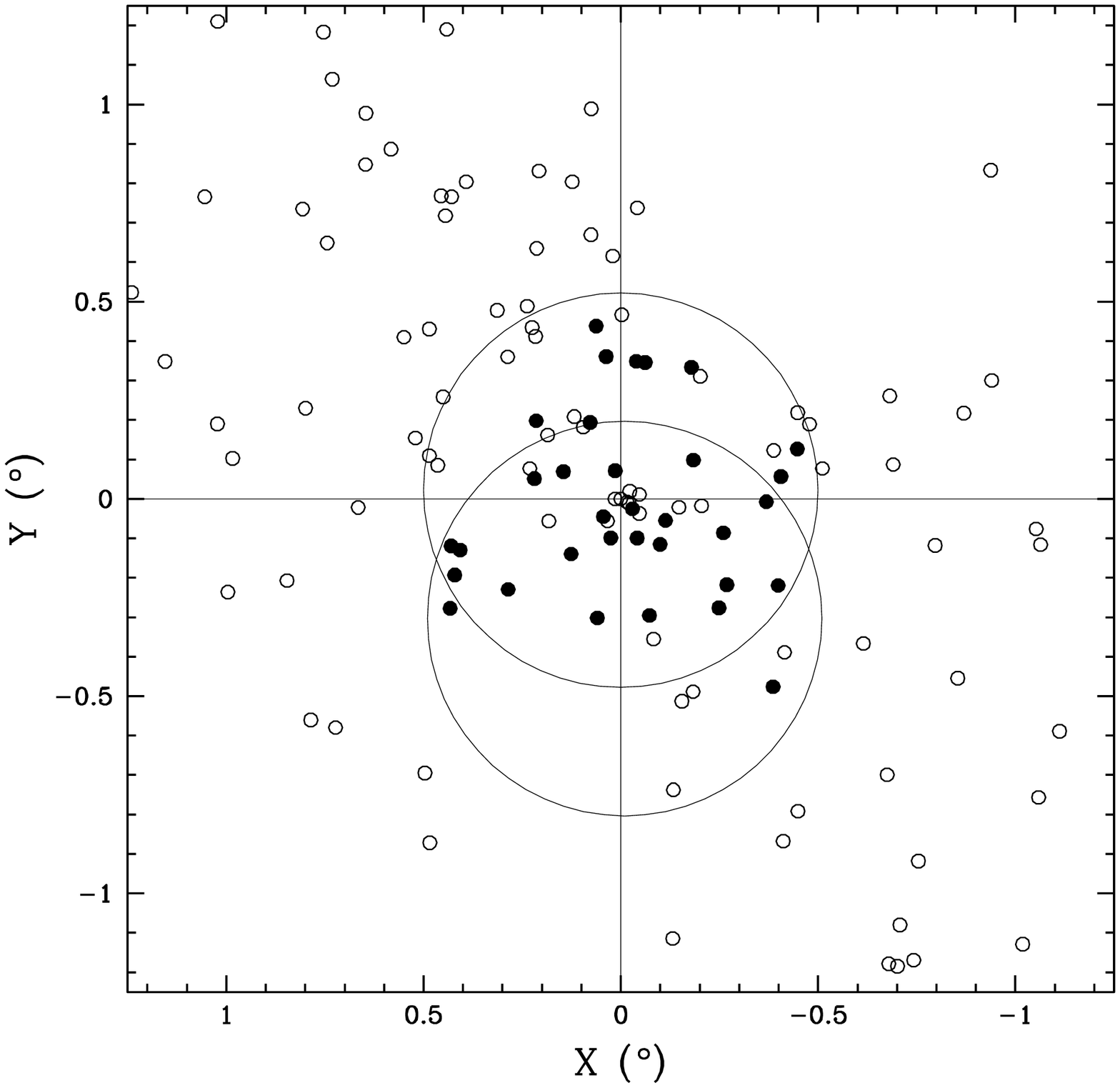}
\figcaption[spat_sam.eps]{
The spatial sampling of A262.  The two pointings are indicated by the large
circles.  The smaller filled circles indicate the galaxies observed by 
AMOS, open circles indicate galaxies with observations from the literature.
\label{spat_sam}}
\end{figure}

A preliminary target list of galaxies brighter than $\about18.5$ in V
within A262 was generated from a 
$1\deg$-square Digitized Sky Survey image centered on the cluster.  Our
goal was to improve the velocity sampling within the virial radius of
the cluster.  We used two AMOS pointings, offset by only 18 arcminutes, to 
improve the density of targets selected within this region.  Final target
selection by the fiber placement algorithm results in a random spatial
sampling of this region.  Combining our observations with previously
published velocities ensures that spatial biasing of the cluster is
minimized.  Figure \ref{spat_sam} shows the spatial distribution of our 
sample and the 
sample of previously published velocities.  The elongation of the cluster
in the SW to NE direction has been seen in every study of A262 (see 
section \ref{applic} for references).

Spectra were obtained using AMOS on the nights of 1998 November 19--20.  
The two pointings yielded 46 object spectra, 29 with an exposure time of 
5400s and 17 with an exposure time of 9000s.  Since the spatial overlap for
these two pointings is significant, the effect of the differing exposure
times on our analysis of the cluster dynamics is negligible.  The 
unassigned fibers monitored the night sky lines during the exposures and 
were used for the night sky subtraction.

All observations were reduced and analyzed using IRAF \citep{tod86}. The
SPECRED package was used to extract one-dimensional spectra from the
registered and coadded CCD images.  A wavelength solution was generated from
calibration lamp spectra and was checked against the night sky lines and
measured to have an internal error of less than 0.5\ang.  We generated a
master sky spectrum for each pointing and used the SKYTWEAK task to scale and
shift the master sky until the RMS of the object spectra was minimized.  We
chose to examine each spectrum individually to eliminate spurious velocities
due to poor sky subtraction.

The redshift of A262 ($z=0.016$) puts the H-$\alpha$ absorption feature in a
region that is relatively free from night sky lines (6671\ang).  For
absorption systems the RVSAO \citep{kur98} task XCSAO was used to
cross-correlate with a zero-velocity K giant spectrum from \citet{jac84}.  For
emission systems the RVSAO task EMSAO was used with a line catalog that
included H-alpha, NII, SII, and OII.  Heliocentric corrections were applied
to derive the final velocities.  As described below, \citet{ton79} R values, 
which measure the quality of the cross-correlations, were one of the
criteria used for removing spurious velocities before combining with
observations from the literature.  The error in the velocity 
is calculated from the width of the cross-correlation peak and the R value 
(see \citet{kur98}) and also contributed to the selection process.

\subsection{Redshifts From the Literature}

The CfA redshift catalog\footnote{available at
http://cfa-www.harvard.edu/$\about$huchra/zcat/} \citep{huc99} formed
the basis of our literature catalog.  A search for more recent data,
not included in the CfA catalog, revealed only the \citet{sco98}
study of early-type galaxies.

Galaxies in common were used to check for systematic offsets between the
three catalogs (CfA, SGH98, and ours) prior to combining them.  Figure
\ref{comp_ab} shows the comparison for the absorption line systems in our
catalog with both literature catalogs. We calculated the average and standard
deviation of the velocity offsets with respect to each of the published
catalogs.  For the CfA catalog the offset is $4\pm259$ \kms\ for
15 galaxies in common.  The offset for the three common galaxies with the 
largest R values is $6\pm81$ \kms. For the SGH98 catalog we derive an offset
of $7\pm56$ \kms\ with 5 galaxies in common.

{
\begin{figure}[p]
\plotone{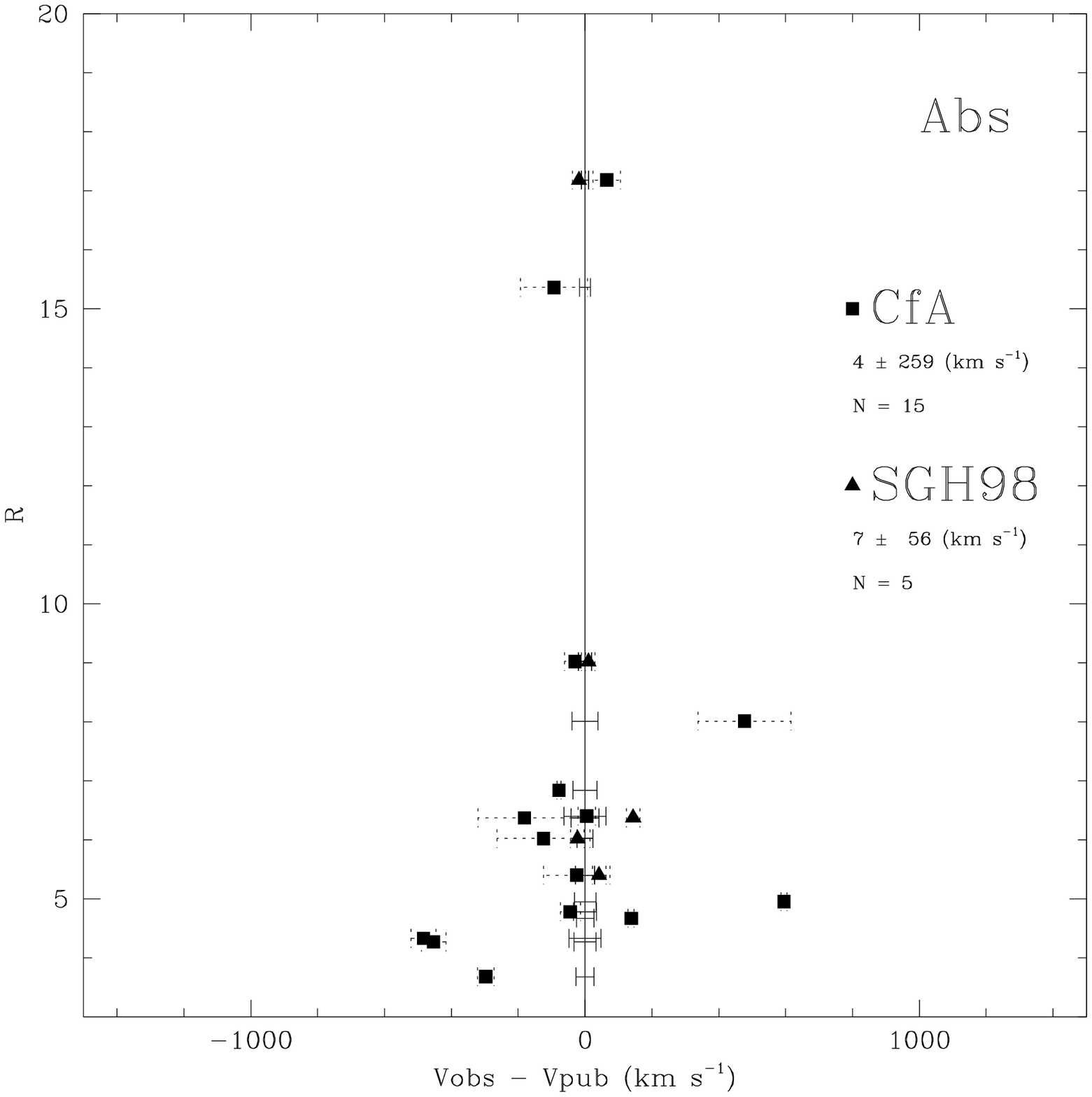}
\figcaption[comp_ab.eps]{
A comparison of our velocities with the CfA redshift catalog and 
SGH98 velocities for absorption line galaxies.  Our error bars are plotted 
along the zero offset line at our measured R value.  Solid points 
indicate the offset from the published value for each previously observed
galaxy with the published error indicated by a dashed error bar.
\label{comp_ab}}
\end{figure}

\begin{figure}[p]
\plotone{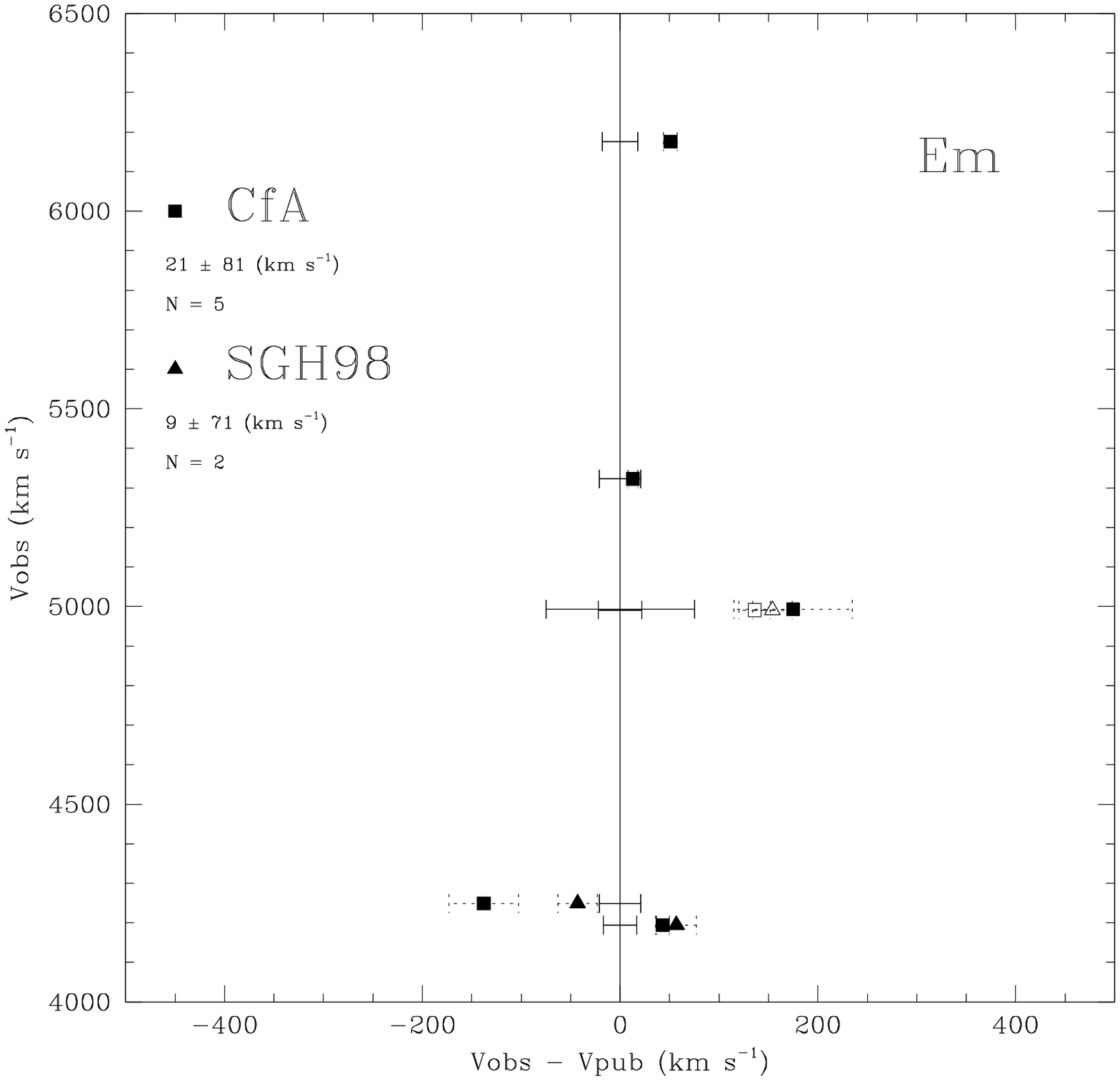}
\figcaption[comp_em.eps]{
A similar comparison to Figure \ref{comp_ab} for the emission line
galaxies, but with the y-axis being our observed velocity.
\label{comp_em}}
\end{figure}
}

We also checked the previously observed emission line galaxies.
This comparison is shown in Figure \ref{comp_em}.  The calculated offset for
the CfA catalog in this case is $22\pm81$ \kms\ for 5 galaxies in 
common.  There are only two galaxies in common with SGH98 
and these yield an offset of $10\pm71$ \kms.

Based on these offsets and standard deviations, we concluded that
there is no statistically significant velocity offset between the
catalogs.  We combined the catalogs into a master catalog of 207 velocities
covering A262 out to a radius of $4\deg$.  The velocities of galaxies with 
multiple observations were combined with a weighted average after checking
each observation for quality.  In the cases where there were three
observations, any velocity that disagreed significantly with two velocities
that agreed was removed before averaging.  In the cases where there were
two observations that disagreed significantly, the good observation was
chosen to have an R value of greater than 5.5 (see Figure \ref{comp_ab}) 
or to have the smaller quoted error.  Table \ref{tabobs} lists the new 
velocities.

The morphologies of our galaxies were derived from their
appearance on the Digitized Sky Survey image.  Most of the brighter 
galaxies had
previously been classified and we used these classifications whenever
possible.  The previously unclassified galaxies in our catalog were divided
into two broad categories depending on the presence or absence of
disk-like structure.  This division was refined using the
spectroscopic appearance by classifying any galaxy with emission lines as a
generic spiral, classifying any disk-like galaxy with no emission lines as an
S0, and classifying the remaining non-disk galaxies with no emission lines as
generic ellipticals.  We checked this classification scheme against the
previously classified galaxies and confirmed its accuracy.

50 galaxies, from 15 different sources within the CfA catalog, had no
published classification.  Of these, 32 have published $B$ magnitudes and
average 0.5 magnitudes fainter than the classified galaxies.  This
unclassified population has interesting properties and 
will be discussed in sections \ref{applic} and \ref{discussion}.

\subsection{X-Ray Data} \label{xray}

We supplemented our optical data with X-ray data from public archives.  We
used {\it ROSAT} HRI data taken on 1996 July 25, {\it ROSAT} PSPC data taken
on 1992 August 10, and {\it ASCA} GIS2 and GIS3 data taken on 1994 January 
22--23.  All datasets were downloaded from the HEASARC data archive.  We used
the HRI data to fit the gas distribution because of its high spatial
resolution, and because of A262's very small core radius (0.01 Mpc $\about$
0.4 arcmin).

We examined the {\it ROSAT} PSPC data to see if we could identify infalling
galaxies by their X-ray emission and remove them from the sample.  Only 4 of
our galaxies had an appreciable count rate in the PSPC image ($>$ 0.01 ct
s$^{-1}$) and removing these from the sample produced no detectable change
in our results.

\begin{figure}[p]
\plotone{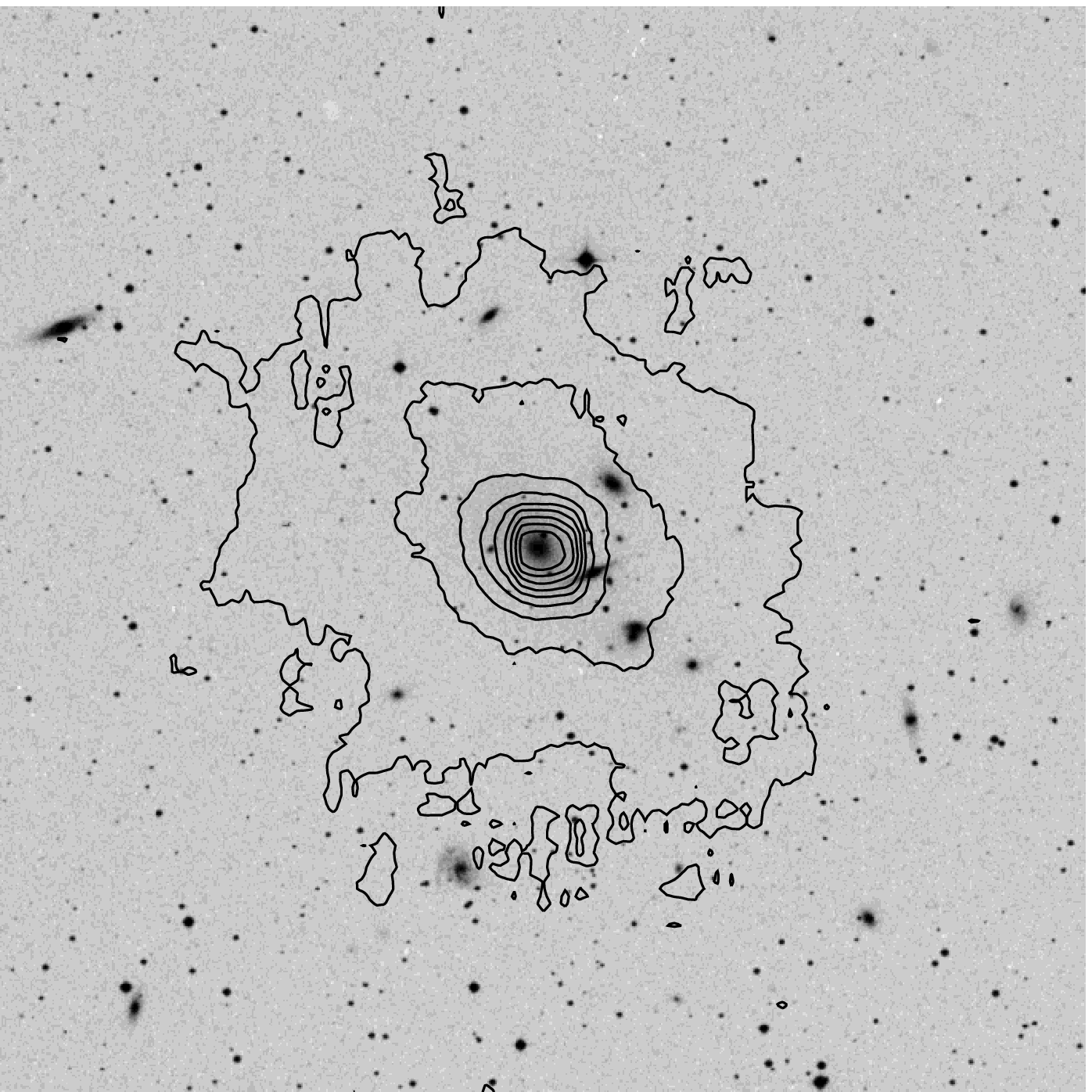}
\figcaption[hri_cont.eps]{
X-ray contours for the {\it ROSAT} HRI overplotted on the DSS image of
A262.
\label{hri_cont}}
\end{figure}

\begin{figure}[p]
\plotone{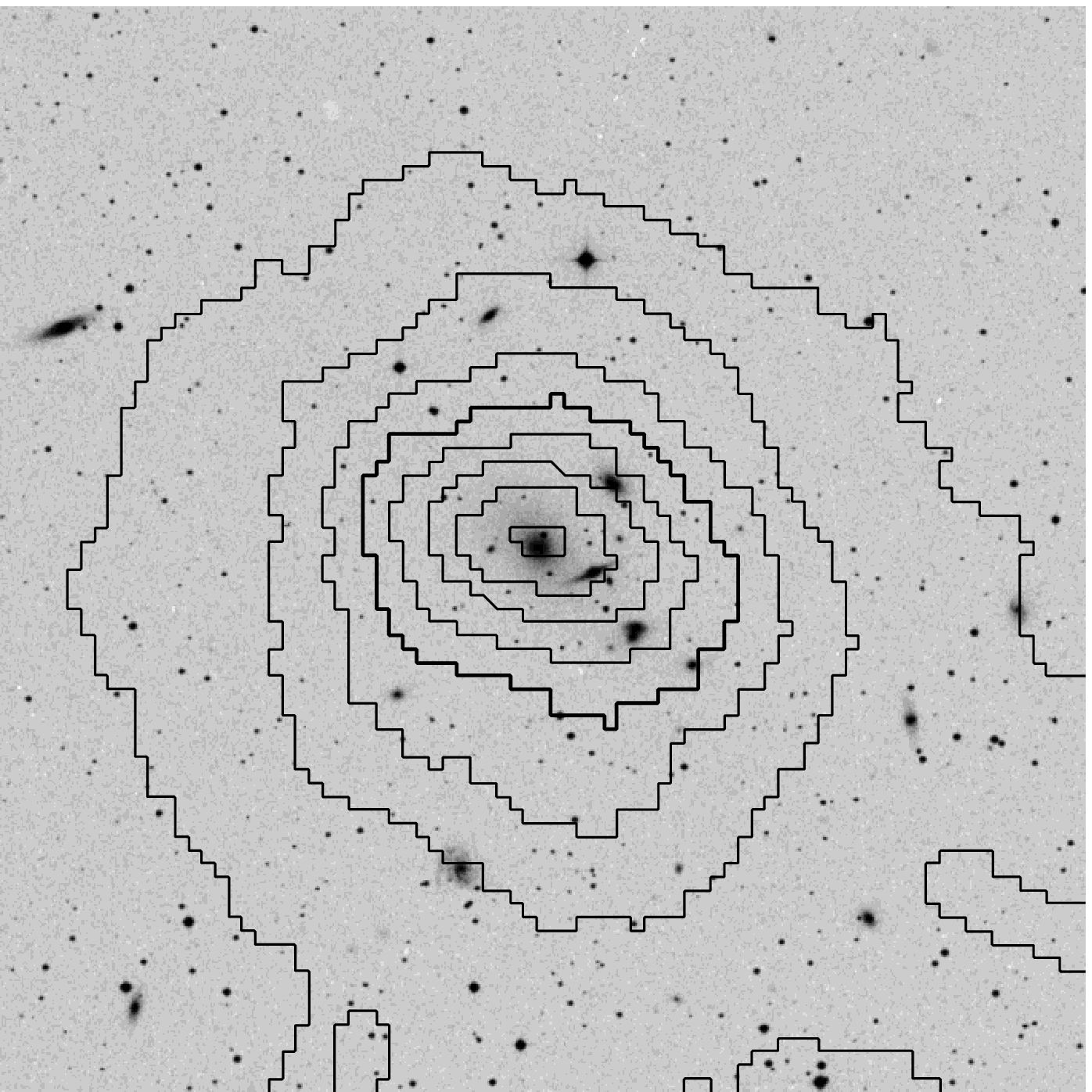}
\figcaption[gis_cont.eps]{
X-ray contours for the {\it ASCA} GIS overplotted on the DSS image of
A262.
\label{gis_cont}}
\end{figure}

Figure \ref{hri_cont} shows the X-ray contours for the {\it ROSAT} HRI and
Figure \ref{gis_cont} shows the X-ray contours for the {\it ASCA GIS} 
overplotted on an image from the Digitized Sky Survey of
A262.  All of the X-ray images show a centrally peaked, slightly elliptical, 
smooth profile.  None show any sign of bi-modality and all have monotonically 
decreasing intensity profiles from the cluster center.  DJF96 analyzed 
the PSPC image
in detail and discuss the correlation between the X-ray and optical
structure.  They found that a modified King profile was consistent with the
profile for A262.

\subsubsection{Profile Fitting}

The X-ray gas profile of A262 was fitted using the vignetting-corrected HRI
image available from the archive.  This image includes data from HRI spectral
bands 2 to 10 and has a total exposure time of 14616s.  We applied the isophote
fitting routine ELLIPSE from IRAF \citep{jed87} and fit from 0--100 pixels
(0--13 arcmin) using the geometric bin spacing algorithm which creates larger
bins for the lower signal wings and smaller bins in the inner region where
the signal is high and the profile is steep.  We then fit the isophotal
values with a modified King model.  \citet{neu99} showed that a point source at
the center of a cluster can systematically bias \bgas and recommend a process 
for fitting the profile that involves removing data points from the 
center until the reduced $\chi^2$ statistic
stops getting smaller.  Others have used a similar technique to remove the
influence of a central cooling flow \citep{dav95}.  Fitting the entire range
of data gives a \bgas of $0.5\pm0.1$ and an X-ray core radius ($r_{cx}$) of
$0.008\pm0.005$ Mpc with a reduced $\chi^2$ of 3.05.  Using the technique of
NA99 we removed the central 5 points (out to 0.2 arcmin) and fit a \bgas of
$0.6\pm0.1$ with an $r_{cx}$ of $0.009\pm0.005$ Mpc at a reduced $\chi^2$ of
1.07.  To test for the influence of the central cD on this profile, we also
fit the data after removing the data points out to the core radius (0.4
arcmin).  This fit yielded values that agreed within the errors.  The
agreement of all these values indicates that the influence of a cooling flow
or a central point source is small.

DJF95 used PSPC data to derive a \bgas of $0.53\pm0.03$ which is consistent
with our value to within the error bars.  The core of A262 would be only
partially resolved in the PSPC and this may have produced a lower \bgas.  Our
$r_{cx}$ also agrees well with the optical core radius ($r_{co}$) measured by
\citet{gir98a}.  G98 measure a value of 0.02 Mpc using a modified King
profile for the surface density of galaxies in the optical.

We adopt a value for \bgas of $0.6\pm0.1$ and $r_{cx}$ of $0.009\pm0.005$ Mpc
for the remainder of our analysis.

\subsubsection{X-ray Temperature}

We utilized the {\it ascascreen} procedure from the HEASARC software archive, 
with the strict
background rejection setting, to filter the raw event files for the two GIS
detector data sets.  The method outlined by \citet{arn96} for
spatial-spectral fitting was followed.  This consists of using XSELECT to
extract spectra in annuli centered on the peak emission.  We divided
the central 12 arcminutes of the image into 4 annuli each 3 arcmin wide.
The temperature gradient of A262 reported in DJF95
is consistent with an isothermal profile and we did not attempt to fit temperatures for the 
individual
annuli.  The corresponding background spectra were extracted from the
standard blank-sky observations covering the same cutoff rigidity range and
with the same selection criteria.  The XSPEC package \citep{arn96} was used
to perform the spectral fitting using the {\it ascac} model which calculates
the energy-dependent scattering between annuli given the spatial distribution
of the emission.  We used our derived profile convolved with the GIS point
spread function for the spatial model.  For the spectral model we used a 
Raymond-Smith
thermal plasma with photo-absorption and allowed the temperature, abundance,
column density and normalization to vary.  This produced a temperature for
the gas of 1.79 KeV (1.71--1.91 KeV 90\% cl) at an abundance of $0.3\pm0.15$
Z$_\odot$ and a
column density of $6.7\pm3.1\times 10^{20}\; {\mathrm cm^2}$ with a reduced
$\chi^2$ of 1.01.

We tested for possible contamination from the higher temperature background
cluster 20 arcmin to the west of A262 (see DJF96).  We performed the same
analysis, first dividing the image into a west half and an east half.  The
derived temperatures for both halves agreed with our previous result 
to within the errors and we
conclude that this background cluster does not affect our analysis.

Our value for the temperature lies between that derived from {\it ROSAT} PSPC
data DJF95; 1.36 KeV (1.21--1.48 KeV 90\% cl), and that derived from
Einstein \citep{dav93}, {\it EXOSAT} \citep{edg91a}; 2.4 KeV (2.2--2.7 KeV
90\% cl), and {\it Ginga} \citep{arn99}; $2.41\pm0.05$ KeV.  The 
{\it Einstein}, {\it EXOSAT}, and {\it Ginga} temperatures may be higher 
because they are contaminated by
the background cluster due to their larger fields of view.
The limited spectral coverage of the {\it ROSAT} PSPC (0.2--2.0 KeV) may have
biased the analysis of those data to lower temperatures.  The {\it ASCA} GIS
has a narrower field of view and good spectral resolution and coverage and
so it does not suffer from either biasing or contamination.  

An X-ray temperature of $1.79^{+0.08}_{-0.05}$ KeV (1$\sigma$) 
is used for the rest of this paper.

\section{DETAILED ANALYSIS} \label{applic}

In this section we present the analysis 
of our redshift catalog to determine
the dynamical properties of A262. 

\subsection{Substructure Tests}

We used our expanded catalog of velocities to identify
the subpopulation of galaxies within A262 that is the most relaxed and
representative of the dark matter potential well.  Other studies have
examined A262 using all populations and have not found
significant substructure \citep{sak94, gir97}.  Girardi et al. used a
multi-scale wavelet analysis with equally weighted spatial and velocity
contributions for a large sample of clusters.  They reproduced 
well-documented substructure in other clusters, but did not detect any in A262.

We find a complex picture emerges when the galaxies are examined by
subpopulation.  The completeness limit of our catalog ($\about16$ in V,
determined using the APS database\footnote{The APS databases are supported by
NASA and the University of Minnesota, and are available at http://aps.umn.edu/})
does not allow us to draw strong conclusions about substructure from the
spatial distribution of the catalog.  However, our conclusions from the
analysis of the velocities should be much more robust since the galaxies 
are then simply test-particles in the potential well of the cluster.  The
distance dimming through the cluster is only 0.1 magnitude so the sampling
through the cluster is uniform.  We first touch on the picture derived from
the spatial distribution of the subpopulations and then examine the
velocity data in detail.

\begin{figure}[p]
\plotone{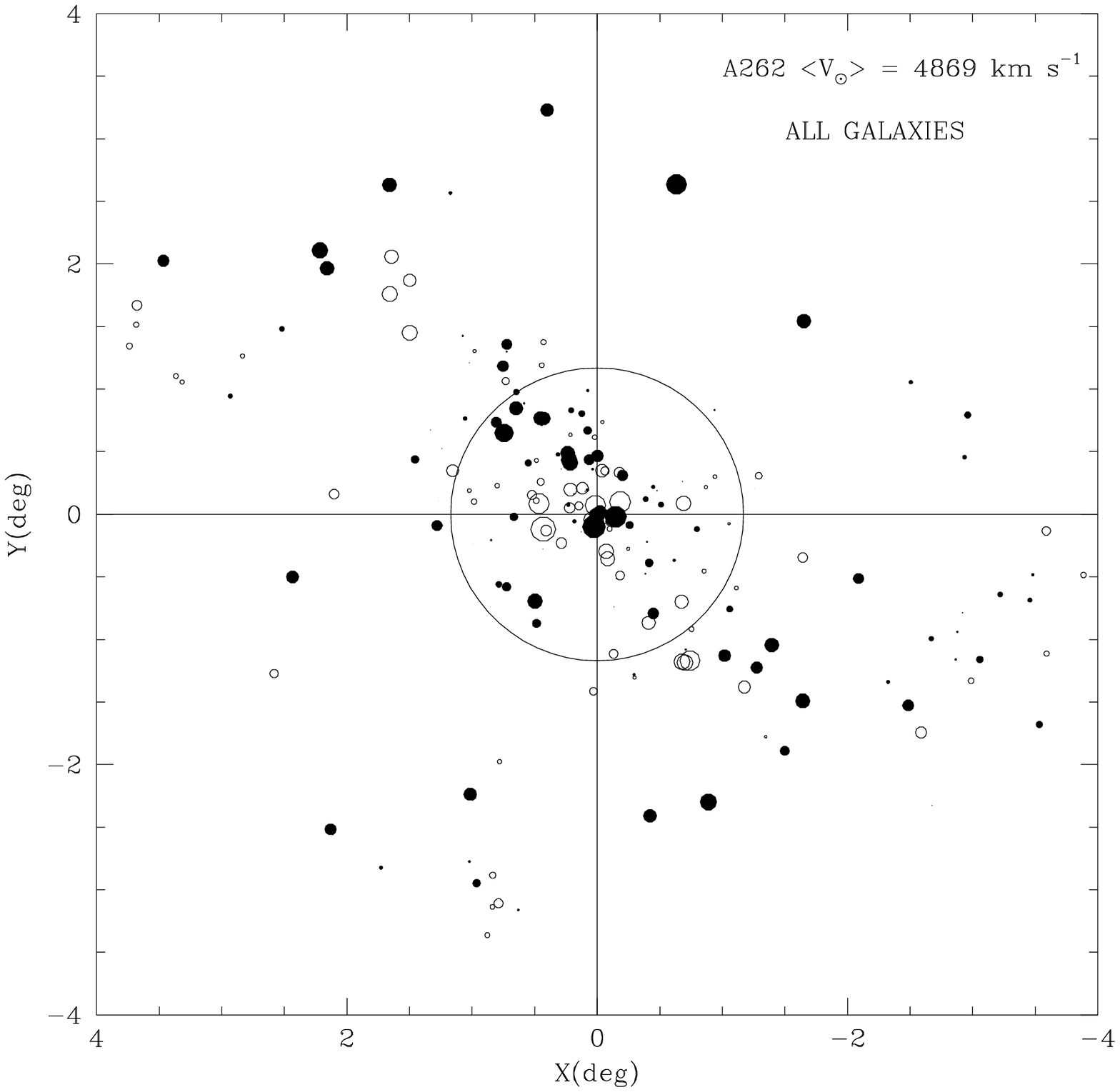}
\figcaption[spat_all.eps]{
The spatial distribution of the entire catalog with peculiar
velocities encoded as follows: filled circles represent
galaxies with negative peculiar velocities, open circles represent galaxies
with positive peculiar velocities, and the size of the circle represents the
magnitude of the peculiar velocity.  The large circle represents the
virial radius of 1.52 Mpc ($1\degpnt17$).
\label{spat_all}}
\end{figure}

\begin{figure}[p]
\plotone{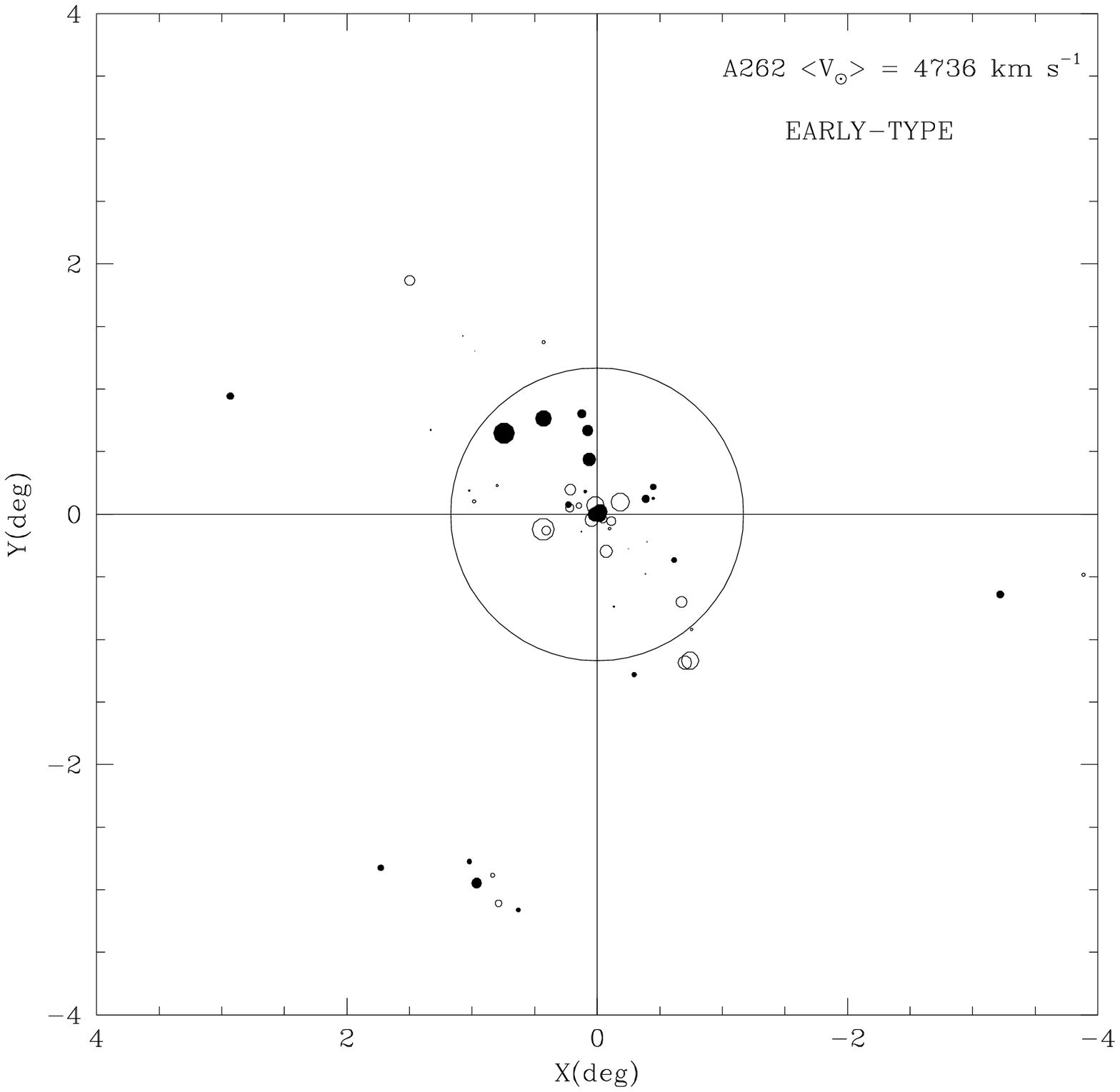}
\figcaption[spat_early.eps]{
As for Figure \ref{spat_all}, but for only the early-type galaxies.
\label{spat_early}}
\end{figure}

\begin{figure}[p]
\plotone{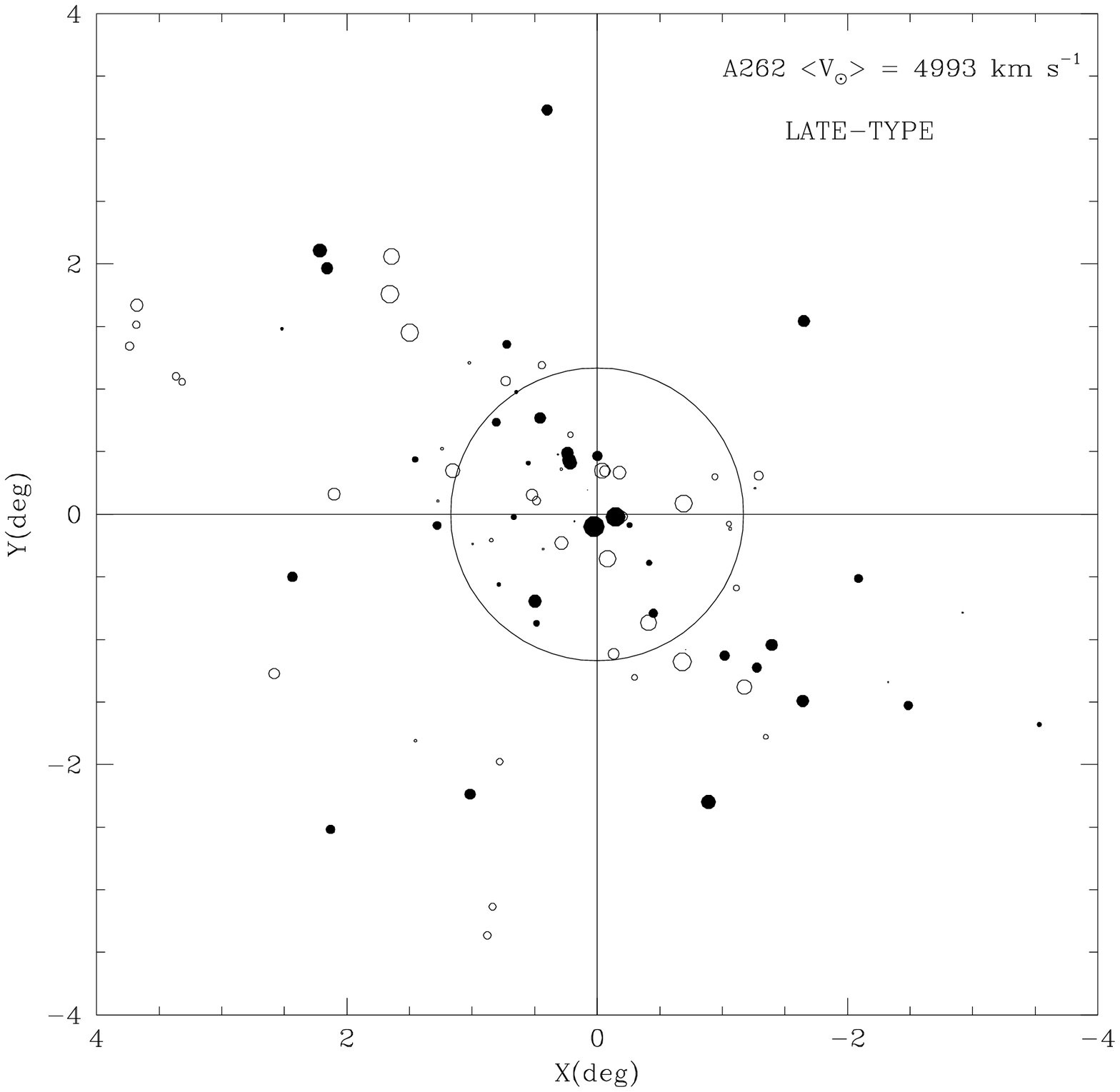}
\figcaption[spat_late.eps]{
As for Figure \ref{spat_all}, but for only the late-type galaxies.
\label{spat_late}}
\end{figure}

Previous studies have noted that A262 has a
centrally-concentrated early-type population of galaxies overlaid with a
later-type population that is infalling from the Pisces-Perseus supercluster
ridge, of which A262 is a member \citep{sak94,mos77}.  Figures 
\ref{spat_all}--\ref{spat_late} display the spatial
distribution of the galaxies with velocities within the
range $3500.0 < V_\odot < 6500.0$ \kms\  and the peculiar velocities with
respect to the cluster average.  Figure \ref{spat_early} shows
that the early-type population is centrally concentrated, while
in Figure \ref{spat_late} the later-type population is clearly more
dispersed. 

\begin{figure}[p]
\plotone{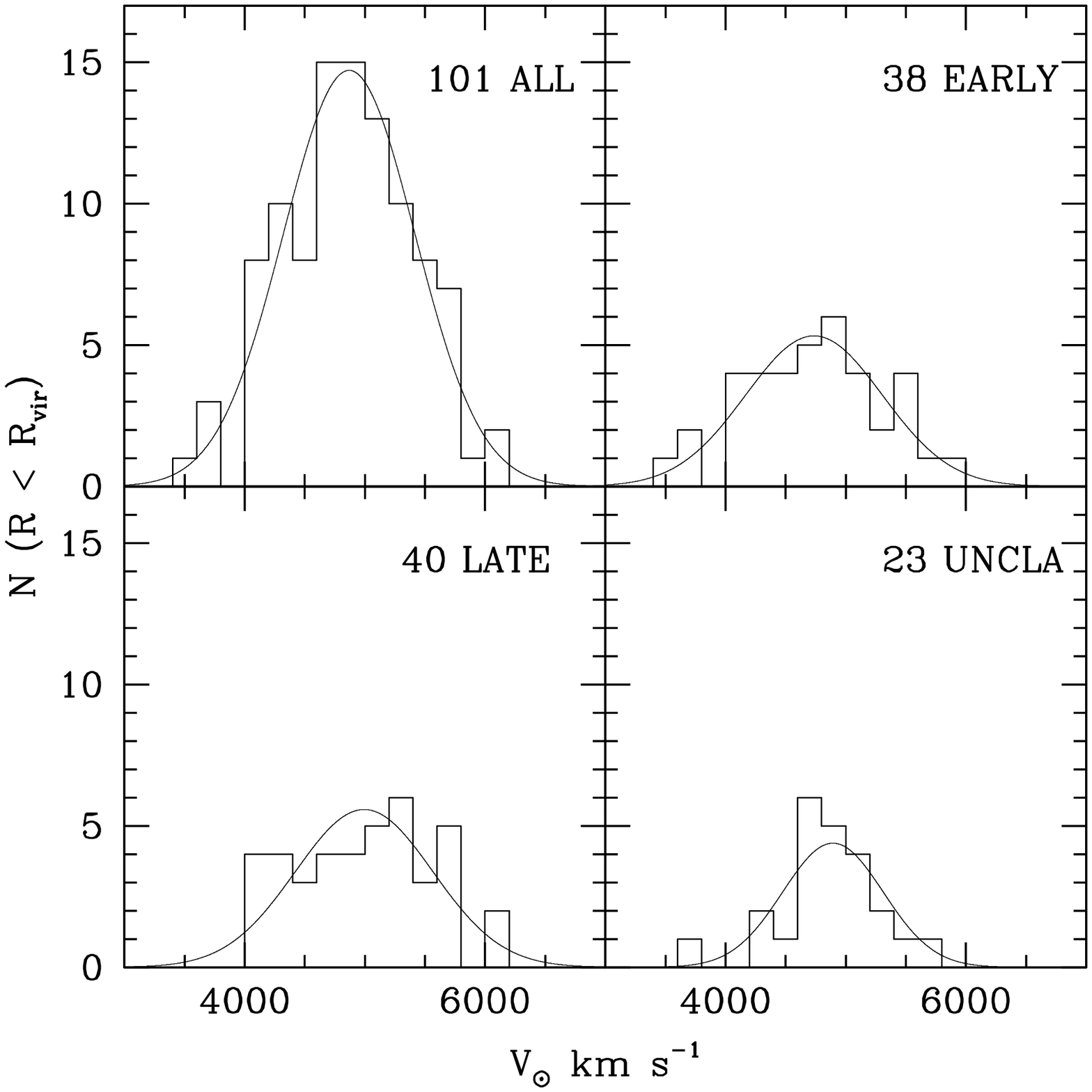}
\figcaption[velhist.eps]{
The velocity histograms for the subsets of the catalog within 
$r_v$ with a Gaussian as specified in Table \ref{tabvel} overplotted.
\label{velhist}}
\end{figure}

We present the velocity histograms for four subpopulations of
galaxies within the virial radius in Figure \ref{velhist}.
We used the ROSTAT algorithm \citep{bee90} to test each of these histograms
for
Gaussianity and to derive the average heliocentric velocity and velocity
dispersion with errors using the bi-weight estimators which are robust to
the presence of outliers.

Table \ref{tabvel} presents the results of this analysis.
The last three columns present the Gaussianity test statistics.  
The first statistic is the scaled tail index (TI, Bird \& Beers 1993)
which varies from 0.84 for a uniform distribution up to a
1.62 for a Cauchy distribution.  A Gaussian distribution has a TI value of
1.0.  The early population is closest to Gaussian using
this statistic, the late population tends toward the uniform distribution,
and the unclassified population is more centrally peaked than a Gaussian,
in agreement with the appearance of the histograms in Figure \ref{velhist}.
The second and third statistics (the a- and W-tests) are described in 
\citet{yah77}.  For the a-test an
infinite Gaussian distributed sample would have the value 0.7979.  Again,
the early population is closest to this value.  The W-test statistic for
a Gaussian has a value of 1.0.  The early population has the value closest
to 1.0 and also has the highest probability of being drawn from a Gaussian
parent population (92\%).

As a test for the effects of incompleteness on our conclusions, we isolated
a complete subsample of our catalog with $V < 16$ and derived the same 
statistics.
The values were identical within the much larger errors (due to the
smaller sample) and we conclude that incompleteness has not effected our
velocity analysis.

\subsection{Average Heliocentric Velocity and Velocity Dispersion Profiles}

\begin{figure}[p]
\plotone{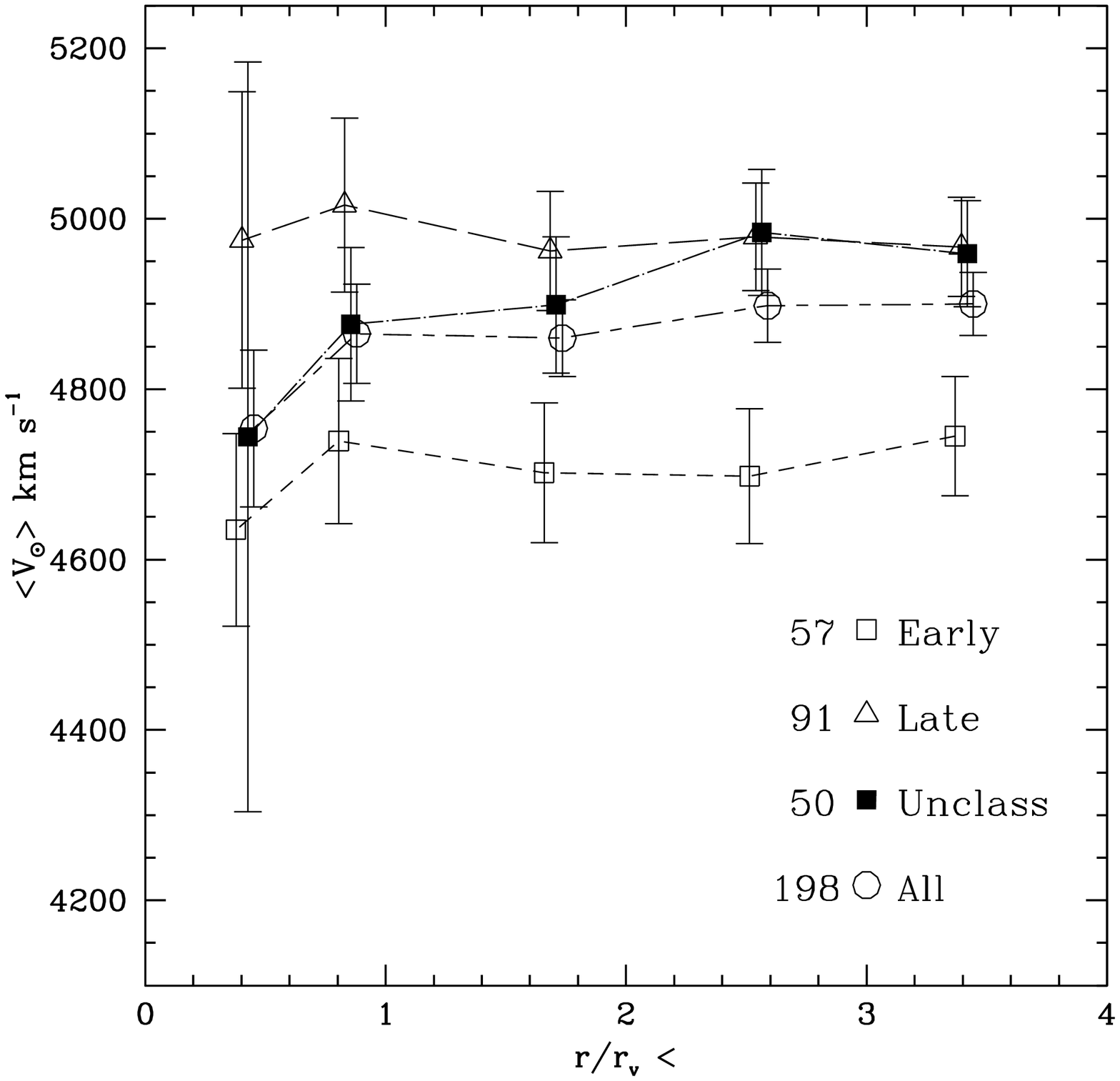}
\figcaption[vhelio.eps]{
The integrated heliocentric velocity profiles for the entire catalog
and several subpopulations. The points are artificially offset 
slightly from each other in $r/r_v$ for clarity.  The lines connecting the 
points are provided as a guide for the eye.
\label{vhelio}}
\end{figure}

\begin{figure}[p]
\plotone{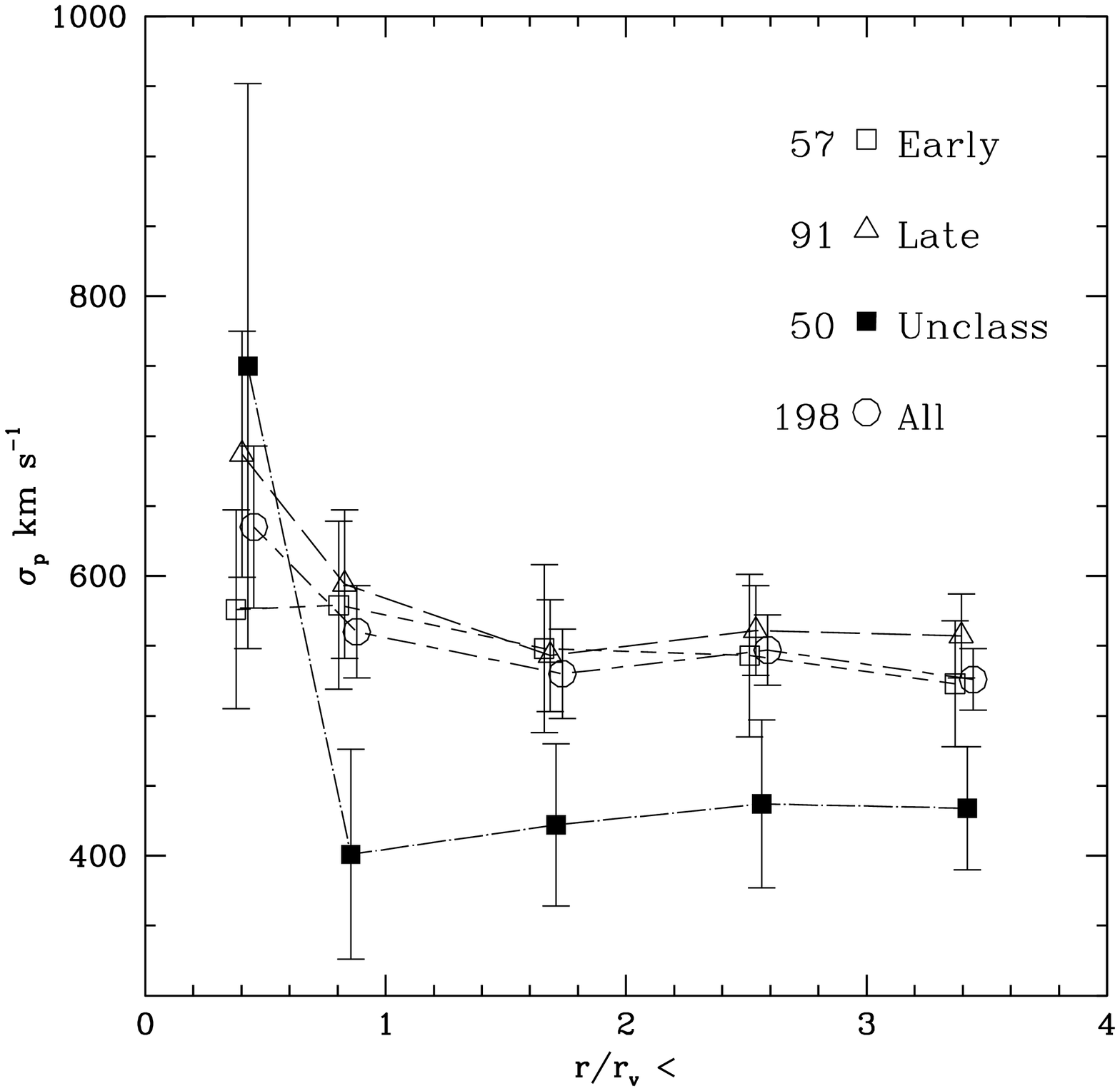}
\figcaption[vdp.eps]{
The integrated velocity dispersion profiles (VDPs) for the whole catalog 
and several subpopulations.
\label{vdp}}
\end{figure}

The average projected integrated heliocentric velocity profiles are 
presented in Figure \ref{vhelio}.  The integrated projected velocity 
dispersion profiles (VDPs) appear in Figure \ref{vdp}.  Each point
was calculated with ROSTAT as described above for each population and 
within the radius indicated.  The cluster center is defined to be the 
position of NGC 708, the central cD.  There is some evidence that NGC 708
may be slightly offset from the dynamical center of the cluster (see
DJF96).  We chose not to account for this due to the small magnitude of the
offset and the large uncertainty in its value.  The
values for the entire cluster population, the early and late-type population,
and the unclassified galaxies are plotted in
the figures.  The error bars are the 1$\sigma$ confidence limits.

Several important features are apparent in these plots.  Each population
shows an isothermal profile both in the VDP and in the heliocentric
velocity.  We tested 2nd order polynomial fits against flat fits using
reduced $\chi^2$ statistics and found that the flat fits did as well or
better than the polynomial fits in each case.  There are two offsets that
are significant in these plots as well.  The early population is offset from
the other populations in
heliocentric velocity and the unclassified population is offset from the
others in the VDP.
We tested the significance of these offsets by using a chi-squared
fitting technique which explicitly accounts for the fact that the true
parameter values for the galaxy populations are unknown, and themselves
estimated from the data.  The best fit value for the early and late
populations have them offset by 260 \kms.  The probability that the early
and late type heliocentric velocity profile data is drawn from a common
parent population is $< \about10^{-5}$.  The same analysis was performed 
to test the reality of the offset of the VDP profile of the unclassified
population versus that of the classified galaxy population.  The
unclassified population VDP is offset from the classified population VDP
by 230 \kms\  and the chance that the two data sets are drawn from a common
parent population is $<\about7\times10^{-4}$.  The VDPs for the late and 
early type galaxy populations are statistically indistinguishable.

\begin{figure}[h]
\plotone{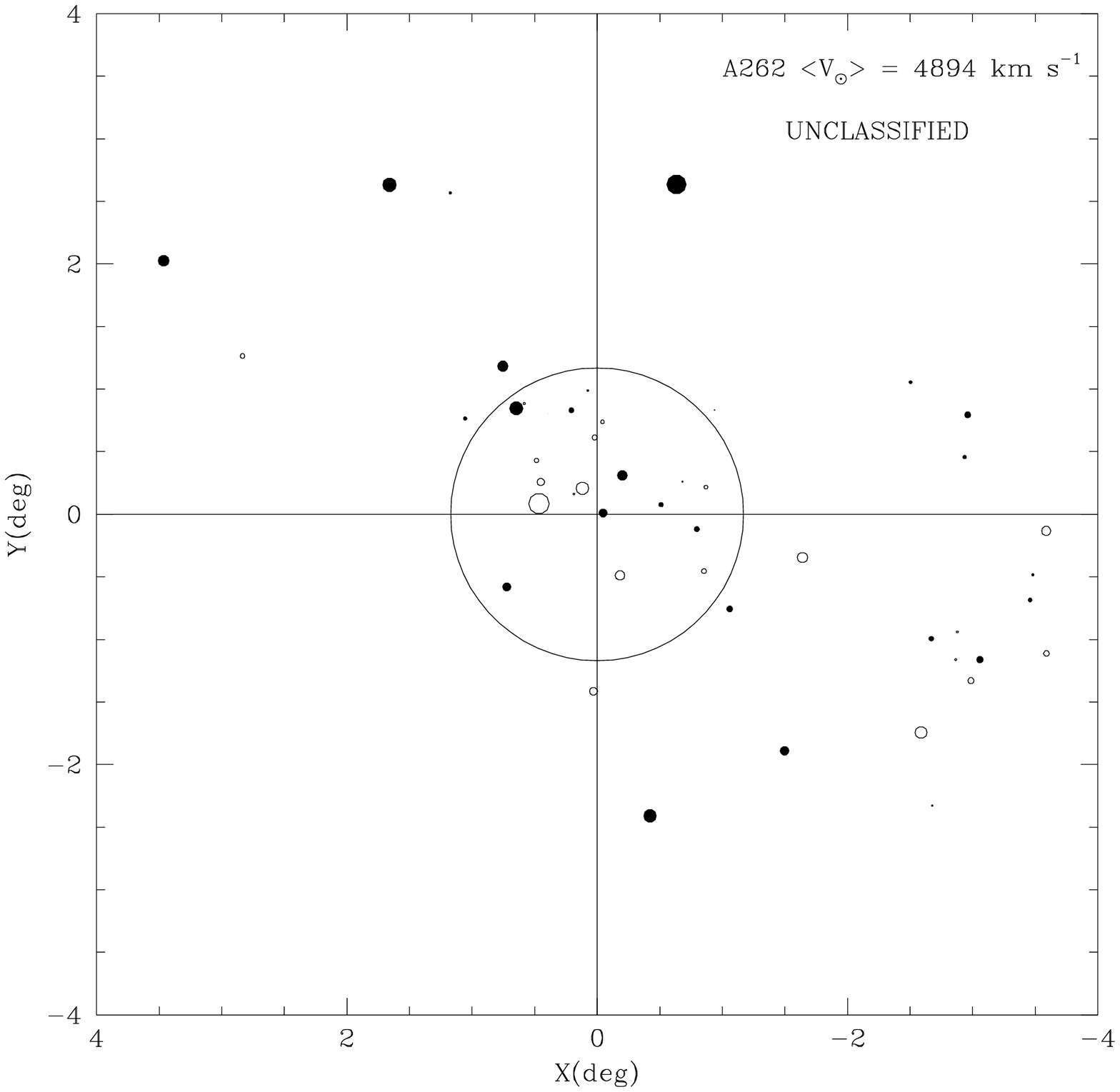}
\figcaption[spat_uncla.eps]{
As for Figure \ref{spat_all}, but for the unclassified galaxies.
\label{spat_uncla}}
\end{figure}

The spatial distribution of the unclassified population is presented in 
Figure \ref{spat_uncla}.  These galaxies cannot be considered relaxed, 
despite their lower velocity dispersion, since they have a very dispersed 
spatial distribution, more like the late-type galaxies in 
Figure \ref{spat_late}.

The profiles and spatial distributions are consistent with the previous
studies of A262 described above.  The isothermal VDP, centrally
concentrated spatial distribution, and Gaussianity tests confirm that
the early-type population is
relaxed.  If infall of the unclassified population occurs along the
supercluster ridge, which is coincident with the plane of the sky, this could
explain their low velocity dispersion.  

Table \ref{tabvel} presents the derived properties out to $r_v$ for the four
population samples of our catalog.  We consider only the early-type
population to be relaxed and will use their values for the remaining
analysis.  It can be seen from Table \ref{tabvel} that the influence of the
late-type population on the velocity dispersion is small because of its
similarity to the early-type velocity dispersion.  The unclassifieds 
lower the velocity dispersion only slightly due to their small number but,
since they are on average 0.5 magnitudes fainter than the the classified 
galaxies, a survey to fainter magnitudes may reveal a greater impact.
We discuss potential explanations for this phenomenon in
Section~\ref{section-unclass}.

\subsection{The Anisotropy Parameter and the Virial Mass}

The equation for determining the radial velocity dispersion and velocity
anisotropy parameter from the projected dispersion is derived in
\citet{bin87}.  Using the notation of DJF96,
\begin{equation}
I(b)\sigp^2(b) = 2 \int_b^\infty \left(1 - A \frac{b^2}{r^2}\right)
                \frac{\rhogal(r)}{\sqrt{r^2 - b^2}}
                \sigr^2 r dr,
\label{eq1}
\end{equation}
where b is the projected radial coordinate in the line-of-sight, \sigp is the
projected velocity dispersion, I(b) is the projected galaxy surface brightness,
\rhogal is the volume density of galaxies, \sigr is the radial velocity dispersion,
and $A$ the velocity anisotropy parameter $A=1-\sigth^2/\sigr^2$.  Using
a modified King profile for the galaxy density of $\rhogal=\rho_0/(1 +
r^2/r_{co}^2)^{3\bgal/2}$, we can solve this equation for the curve
$\sigr(A)$ that reproduces the measured \sigp. 
In practice, due to 
observational errors (primarily from \sigp), we
find a region in the ($\sigr,A$) plane that satisfies the constraint.
The optical core radius of the cluster galaxies is $r_{co}$. The core radius
we used for the King profile was $r_{co}$ determined in G98. The influence of a
central cD galaxy can complicate the determination of this quantity, but the
solution region is fairly insensitive to its value.
We note that DJF96 simply used the $r_{cx}$ they determined from the
{\it ROSAT} PSPC data, assuming that $r_{co}$ and $r_{cx}$ were equal.

We assumed that $A$ and \sigr are independent of radius
in the cluster. In order to further limit the solution set we impose a second
constraint on \sigr and $A$ which comes from the virial mass.  The connection
is in the form of the Jeans equation \citep{bin87}
\begin{equation}
M(r) = \frac{-\sigr^2 r}{G}
        \left( \frac{d{\mathrm ln}\rhogal}{d{\mathrm ln}r} +
        \frac{d{\mathrm ln}\sigr^2}{d{\mathrm ln}r} + 2A\right).
\label{eq2}
\end{equation}
DJF96 deduced \sigr and $A$ using the X-ray determined virial mass in equation
(\ref{eq2}).  Recently G98 has shown that excellent agreement can be obtained
between X-ray and optical virial masses provided the optical virial masses
are corrected for a surface energy term.  This term accounts for the fact
that observations are
made out to a finite radius in the cluster. This correction has been
emphasized in work on CNOC clusters \citep{car96}.  The corrected optical
virial mass can be expressed in terms of the standard virial mass according
to the following equation (G98):
\begin{equation}
M_{cv} = M_v \left( 1 -
        \frac{4 \pi b^3 \rhogal(b)}{\int_0^b 4 \pi r^2 \rhogal(r) dr}\;
        \left[ \frac{\sigr(b)}{\sigma(<b)} \right]^2 \right),
\label{eq3}
\end{equation}
where b is the boundary radius, $\rhogal(b)$ the galaxy density evaluated at
the boundary radius, $\sigr(b)$ the radial optical dispersion at the boundary
radius, $\sigma^2(<b)$ the total optical dispersion ($=\sigr^2 + \sigth^2 +
\sigma_\phi^2$) averaged out to the boundary radius.  The uncorrected
(or standard) virial mass is given by
\begin{equation}
        M_v = \frac{3 \sigp^2 r_v}{G},
\label{eq3a}
\end{equation}
(G98, eq. [4]).  Given the virial mass,
equations (\ref{eq1}) and (\ref{eq2}) further constrain the solution for
{($\sigr,A$).}

The values of ($\sigr,A$) so obtained will also provide a more accurate
corrected virial mass.  In quoting the virial mass below we refer to the
total virial mass.  As discussed by \citet{car96} this is
the mass contained within a radius of virialization defined by
\begin{equation}
\bar{\rho}(r_v)/\rhoc(z) = \frac{6}{(1 + z)^2 (1 + \omega0 z)}\;
        \frac{\sigp^2}{\H0^2r_v^2},
\label{eq4}
\end{equation}
where \rhoc is the critical mass density at redshift $z$ and
$\bar{\rho}(r_v)$ the mean density of the cluster out to the virial radius.
Essentially all the virialized mass is accounted for if
$\bar{\rho}(r_v)/\rhoc(z)=200$ \citep{cro94,gun72}.  G98 give an
approximate method for determining the radius of virialization and 
\citet{gir98b} give more exact iterative methods for determining the radius of
virialization for different cosmological models.  The virial mass we quote
below is based on equation (\ref{eq4}) which gives $r_v=1.52$ Mpc 
($1\degpnt17$) and, for the low redshift of A262, differs by 
$ <\about15\%$ from the $r_v$ quoted in G98.

\begin{figure}[h]
\plotone{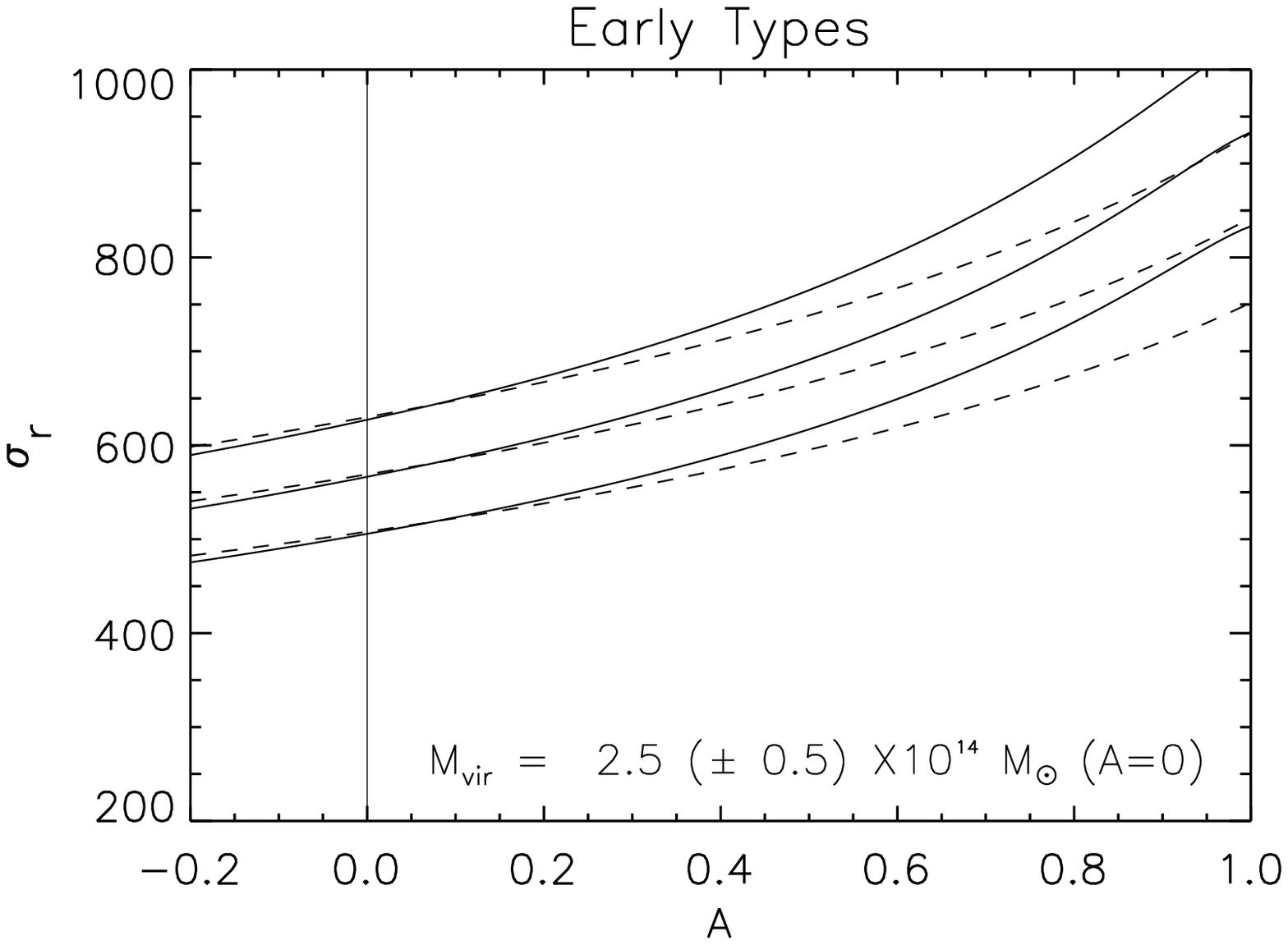}
\figcaption[ani_early.eps]{
The range in radial velocity dispersion, \sigr, and anisotropy
parameter, $A$, that reproduce the projected velocity dispersion, \sigp,
are shown with dashed lines.  Solid lines show the range in \sigr and $A$
that reproduce the corrected virial mass, $M_{cv}$, from the stellar
hydrodynamics equation (eq. [\ref{eq2}]).
\label{ani_early}}
\end{figure}

We used the following method to determine the range of orbital anisotropy, $A$, and
radial velocity dispersion, \sigr, that produce consistency between equations
(\ref{eq1}) and (\ref{eq2}).  We used $\bgal=0.73$ from G98.  We
calculated the standard virial mass from equation (\ref{eq3a}) using the
early-type \sigp.  Then, from equation (\ref{eq1}) we get a range of \sigr and
$A$.  We use these values to calculate the corrected virial mass from
equation (\ref{eq3}) as a function of $A$ and use this in equation (\ref{eq2}). 
Figure \ref{ani_early} plots the locus of \sigr, and $A$ from equation 
(\ref{eq1}) (dashed lines) and from equation (\ref{eq2}) (solid lines) for 
the early-type galaxies in A262.

Unlike DJF96, we get complete overlap between these two distributions over
all reasonable values of $A$ (see below).  We get a standard virial mass of
4.2$\times$10$^{14}$ M$_\odot$ which is slightly higher than 
the value in G98 due
to our higher \sigp, however our corrected virial mass (2.5$\times$10$^{14}$
M$_\odot$) is greater by $\about$25\% (at $A=0$).  G98 used a statistical
method for determining the value of $A$ used in computing the mass
correction.  They divided all VDPs into three categories: centrally rising,
flat, and centrally falling.  They assigned A262 to the centrally rising
category.  With our larger catalog, we measure a
flat VDP for the global population as well as each subpopulation.  This
accounts for the difference in our correction to the virial mass.

This analysis does not strongly constrain the range of A.  In the
following analysis we will take $A$ to be zero, but will also consider the
implications of a range of velocity anisotropies for the determination of
\bspec.

\section{DISCUSSION} \label{discussion}

\subsection{The $\beta$--problem}

The cluster gas is assumed to be described by the equation of hydrostatic
equilibrium \citep{cav76,fab80} and this leads to an expression for
the cluster mass
\begin{equation}
M(r) = \frac{-kTr}{G \mu m_p}
        \left( \frac{d{\mathrm ln}\rhogas}{d{\mathrm ln}r} +
        \frac{d{\mathrm ln}T}{d{\mathrm ln}r} \right),
\label{eq5}
\end{equation}
where $\mu m_p$ is the mean atomic weight of the gas and $T$ the intra-cluster
gas temperature.  Equating the masses in equations (\ref{eq2}) and (\ref{eq5}) leads to an
expression for the ratio of galaxy radial velocity dispersion to gas
temperature \citep{bah94}.  Using the notation of G98 with the modified King
profiles and performing the logarithmic derivatives we obtain
\begin{equation}
        \bspec \equiv \frac{\sigr^2}{kT/\mu m_p},
        \label{eq6a}
\end{equation}
\begin{equation}
        \bfitc \equiv \frac{\bgas/\bgal}{1 - 2A/{3 \bgal}},
        \label{eq6b}
\end{equation}
and
\begin{equation}
        \bspec = \bfitc.
        \label{eq6c}
\end{equation}
In the case where the gas is isothermal and where the anisotropy parameter
$A$ is negligible we can identify \bspec with the ratio of cluster galaxy
internal energy to gas kinetic energy.  In this case \bfitc just reduces to
$\bfitc=\bfit\equiv\bgas/\bgal$.  The failure of \bspec to equal \bfitc
is the $\beta$--problem.

Using $\sigp=569\pm61$ \kms, T$_x=1.79^{+0.08}_{-0.05}$ KeV, and assuming
$A=0$ we get a \bspec of $1.13\pm0.12$.  With $\bgas=0.6\pm0.1$ and
\bgal from G98 of 0.73 we get a \bfitc of $0.82\pm0.20$.  
Our analysis obtains values for \bspec and \bfitc which are
consistent to within statistical uncertainties; there is no $\beta$--problem.
The \bspec obtained in this study is twice as high as that
previously obtained (see Table \ref{tabbetas}).  The \bfit we report is 
$\about50\%$ higher than prior determinations due to a new, lower \bgal 
for A262 reported in G98.  
As a check on the G98 value, we used our complete galaxy sample to calculate
\bgal.  Our \bgal is statistically indistinguishable from the G98 value, but
has a larger uncertainty.  For the early type
population we obtain a slightly higher \bgal but the errors are
even larger for this very limited sample and our result is, within errors,
consistent with G98.  Therefore we have adopted the G98 number as the
best current value of \bgal.

Our conclusion that there is no $\beta$--problem in A262, while the same as
that drawn by DJF96, is based on very different observational inputs
and contains considerable subtleties.  Moreover, although we both conclude
that there is no $\beta$--problem, our higher \bspec has different implications
(discussed below) for galaxy-gas feedback models.  It is probably
correct to discount the late-type galaxies in calculating \bspec but
not because they have a different velocity dispersion from the early-type 
galaxies, as DJF96 suggested.  Indeed, the late-type galaxies'
velocity dispersion and VDP are statistically
indistinguishable from the early-type galaxies and thus they do not
affect the calculation of \bspec at all.  But, they clearly represent
a distinct, unrelaxed cluster population.  This manifests itself in
their heliocentric velocity which differs from the early-type galaxies,
their spatially extended distribution, and their non-Gaussian (but
isothermal) velocity distribution. The unclassified population has an
isothermal VDP but does have a statistically significant difference in
velocity dispersion (see section \ref{section-unclass}).
This population also has a peculiar heliocentric velocity. They 
should definitely
be excluded from any calculation of \bspec in A262, but as a
practical matter they are a small enough population (at least in A262)
that they do not seriously affect the value of \bspec . 
Finally, the isothermal temperature profile of the galaxies, and lack of a
significant X-ray temperature gradient (reported in DJF95)
all argue against anomalies that might alter the conclusion that the 
$\beta$--problem does not exist in this cluster.

The basis for the difference in \bspec between this work and 
DJF96 is twofold.  Firstly,
our measurement of the velocity dispersion 
of the virialized early-type galaxies (569 \kms) is
much higher than the 330 \kms\  \citep{mos77} used by
DJF96.  The low value of velocity dispersion was based on only 9
elliptical galaxies, which shows the importance of obtaining a
statistically large sample for analysis of subpopulations within a
cluster. Secondly, the X-ray temperature we determined from the ASCA
data is higher than that determined by DJF96 from the
ROSAT data. It is, however, the large difference
in velocity dispersion between our samples that dominates the
difference in \bspec.

\subsubsection{Implications for the Dynamics of A262}

The solution to the $\beta$--problem presented here, at twice the previously
reported {\bspec} , has implications for the general state of A262.  We have
quoted a \bspec solution for the assumption that the velocity anisotropy
parameter $A=0$. The flat VDPs that we measure for the early-type galaxy
population suggest $A=0$ is a plausible supposition.  However, as Figure
\ref{ani_early} indicates, the radial velocity and $A$ are not
well-constrained by either the Jeans equation method or the line-of-sight
velocity dispersion method.  Indeed, these two approaches produce virtually
identical constraints on radial velocity and $A$. The two methods should give
the same constraints but often do not.  Our excellent agreement results
from using the corrected optical virial mass as determined by the method of
\citet{car96} (Equation \ref{eq3}) in the Jeans equation and then iterating
to demand self-consistency between the Jeans equation and the corrected
virial mass equation. With this procedure the Jeans equation is determining
the correct \sigr and $A$ relation, just as the line-of-sight velocity
dispersion method does (Equation \ref{eq1}).  Indeed, it is precisely because
we are properly utilizing the Jeans equation that it provides no extra
information about the relation between \sigr and $A$ beyond that obtained
from the line-of-sight velocity dispersion equation.  However, the excellent
agreement between these determinations does tell us that the early type
galaxy population used in the analysis is well-virialized.

Aside from assuming the plausible $A=0$, we can draw a
coarse conclusion about the maximum value of $A$ in this cluster. The
very largest values of \bspec , measured for clusters in which there is minimal
substructure, is $\about1.3$ (BMM95).  
For our nominal curve of $\sigr(A)$, this would
imply a rough upper limit of $A < \about0.25$.  This is entirely consistent
with the work of \citet{van99} who found a range of $A$
from $-0.05$ to $0.26$ for their cluster ensemble derived from 16 clusters
in the CNOC1 survey.

The higher \bspec we have found for this cluster is most relevant to
discussions of galaxy-gas feedback and entropy-floor models. If $A=0$,
then our \bspec for A262 is consistent with the mean value of
\bspec ($0.91$, see Table \ref{tabbetas}) determined for several 
large cluster
samples which either had substructure corrections applied or involved
luminous X-ray selected systems claimed to be less likely to have 
substructure.  A
larger \bspec has been obtained from a cluster sample of more
heterogeneous morphology ($1.14$, see Table \ref{tabbetas}). Our 90\%
lower limit on
\bspec is 0.95 for $A=0$.  Assuming that $A=-0.05$, the lowest $A$
obtained in the van der Marel et. al. sample, the lower limit on \bspec
is $0.92$.  These lower limits both imply that A262 could have at
most a $\about5\%$ enhancement of gas temperature with respect to galaxy
temperature.

This (at most) mild overheating of gas with respect to galaxies we
obtain for A262 is in marked contrast to the factor of 2 overheating
inferred from the \bspec of DJF96.  It is interesting to place our
results in the context of models for galaxy-gas feedback through winds in
early galaxies or through dynamical friction in galaxies.  Several
authors have suggested a correlation between cluster velocity
dispersion and gas temperature steeper than predicted by the virial
equation (see BMM95 and references therein) but the scatter in the
cluster data is large.  Nevertheless, a steeper relation between
velocity dispersion and gas temperature (effectively 
a temperature-dependent \bspec) can be argued as resulting from winds injecting
energy into the cluster or from dynamic friction of galaxies in the
cluster.  

The simulations of \citet{met94} as presented in
BMM95, which include the dark matter, galaxies and winds, offer
support for the contention that \bspec is temperature dependent.
Their simulations would predict, given the temperature of A262, that
\bspec $\about0.6$.  This is substantially lower than the lower limit we have
derived for \bspec .  The simulations suggest that a cluster would
need a $T > \about3$ KeV to obtain the type of \bspec we measure.

There are possible problems with this analysis.  As pointed out by
BMM95, to obtain this T-dependent \bspec , the simulations required the
use of much greater wind luminosities for early galaxies than are
realistic. The assumption that galaxies can be modeled as
collisionless particles is also questionable \citep{fre96}.
Since A262 gives results consistent in all respects with a
well-relaxed system suffering 
from minimal substructure complications in the
early-type galaxy population, the discrepancy with the predictions of
wind models are hard to ignore. While the analysis in BMM95 
lends support to the idea of a T-dependent \bspec, it is not clear on the
whole whether that translates into support for the galaxy wind models.
Clearly, further work is warranted on models attempting to predict this
velocity dispersion-temperature correlation.

We have demonstrated there is no $\beta$--problem in A262.  Our solution is
consistent with \citet{edg91b}, who suggested the $\beta$--problem
is likely associated with improper determinations of velocity dispersion,
as emphasized in BMM95.  Alternately, it should be pointed out that
Bahcall and Lubin (1994) have proposed that the beta-problem vanishes in
general when one utilizes the proper average net
galaxy density around clusters, as determined from galaxy-cluster
cross-correlation.  This approach provides a
shallower dependence on \rhogas ($\rhogas\;\prop\;r^{-2.2}$, Schombert
1988; Bahcall 1977) than the canonical King approximation
($\rhogas\;\prop\;r^{-3}$).  It produces a value of \bfit which is in
better agreement with \bspec when both are suitably averaged over an 
ensemble of clusters.  The agreement is purported to hold even in the case of
non-isothermal clusters (cf. Eq. \ref{eq5}).
This is not in agreement with BMM95, since their solution requires only
gravitational energy to feed the cluster processes.  It is possible
that all these works suffer from the shortcoming of relying on
ensembles of clusters to draw conclusions about \bspec.  Additional
insights should be obtained as more individual clusters are scrutinized
with detailed optical and X-ray studies to form a comprehensive picture
of their interactions.

\subsubsection{\bspec, The Unclassified Population and Velocity
Fields}\label{section-unclass}

Two unvirialized galaxy subpopulations have been identified in A262.
The late-type galaxies have a mean velocity dispersion averaged over the
virial radius and a VDP which are statistically the same as those for the 
early-type galaxies.  They are dynamically distinct in that they show a
much less concentrated spatial distribution and have a significant velocity
offset with respect to the early-type galaxies.  The unclassified galaxies
are distinct from the early-type galaxies for these same two reasons, but in
addition have a significantly lower (but flat) VDP.  Based on their
appearance in Figure \ref{vhelio}, the unclassifieds are a mix
of early and late-type galaxies in the central bin and become more
dominated by the late-type galaxies as $r$ increases.

The most peculiar aspect of the unclassified galaxies is their low VDP.
If they represent a more recent infalling wave of galaxies whose peculiar
motion is along the plane of the sky, this would explain their lower velocity
dispersion.  They could also be a `local field' population originating from the
Pisces-Perseus supercluster ridge which is coincident with the plane of the
sky. These
low-luminosity and presumably low-mass galaxies could have been ejected from
other clusters in the supercluster ridge through mass segregation.  In this
description, the early-type galaxies trace the cluster potential well, the
late-type (classified) galaxies have infallen, but not yet equilibrated with
the cluster potential, and the unclassified galaxies trace the background
supercluster at a constant velocity offset and lower velocity dispersion.
\citet{edg91b} have previously discussed how such field galaxies, 
if not properly 
accounted for, may contribute to the existence of a $\beta$--problem.

\subsubsection{The Virial Mass of A262}

Using the optical and X-ray analyses we can compare the virial masses
determined by the two methods.  The equation of hydrostatic
equilibrium (see DJF95, Eq. 3) yields an X-ray mass of 
$M_x = 1.9 \pm0.3 \times 10^{14} M_\odot$.  From our velocity dispersion
within the virial radius, assuming an anisotropy parameter of $A=0$
and using the correction of \citet{car96}, 
we derive an optical mass of $M_{opt} = 2.5 \pm 0.5 \times 10^{14} M_\odot$.
The agreement between
the two determinations is quite good (25\%) and 
well within the statistical errors.
G98 reported that optical and X-ray determined
virial masses for a large ensemble of clusters could on average be
brought into a statistical agreement of $\about30\%$ if the optical virial masses
were corrected for the surface term in the virial equation as
prescribed in \citet{car96}.  We have shown here that the same
techniques, employed on individual clusters in which a virialized
galaxy subpopulation can be identified, also yields good agreement
between X-ray and optical masses.

\subsection{Entropy Floor Models, and the $L_x-T$ Relation}

DJF96 concluded that the gas temperature in A262 was a factor of 2
higher than the galaxy temperature.  A high (but not that high) gas
temperature relative to the galaxy temperature fits comfortably into a larger
theoretical framework for understanding clusters.

Observations of low richness class clusters, including A262, show that the
cluster entropy, when plotted as a function of radius normalized to the
cluster virial radius, tends to be flat in the interior regions, and
increases sharply at larger radii.  This is roughly consistent with
predictions of the ``entropy floor'' model (Evrard \& Henry 1991; DJF96).  
In the
entropy floor model there is assumed to be substantial preheating of the
proto-intracluster medium at high redshift. This heating leads to an
adiabatic cluster inner region, and increasing entropy in the outer
regions due to shock heating of infalling gas.  The entropy floor model not
only makes a specific prediction about the entropy as a function of cluster
radius (normalized to the virial radius), but it also makes a definite
prediction about the cluster $L_x-T$ relation.  Recent observational data on
several sets of clusters \citep{hen97,arn99} suggest an $L_x-T$
relation for the entropy floor model which is roughly consistent with the
observed slope of the $L_x-T$ relation.  The entropy floor model also
correctly predicts the observed minimal (or no) evolution of the $L_x-T$
relation with redshift.  More complex models also make predictions
which are in accord with current data, including semi-analytic models in
which the $L_x-T$ slope changes with cluster temperature \citep{cav97}, the
punctuated equilibrium model \citep{cav99} and certain wind models
\citep{met94}.

The entropy floor model does appear to provide agreement with the
radial entropy profile in A262 as determined by DJF96.  The flatter gas
profile is also in accord with models in which there is early heating
of the gas.  These models also predict substantially higher gas than
galaxy temperatures in poor richness class clusters.  The higher gas
temperature is expected because energy input from the galaxies into the
gas is roughly independent of the richness class, leading to
disproportionate heating in the low richness class clusters.  Our
\bspec range in A262 is consistent with no overheating of gas with
respect to galaxies, or a very mild overheating ($<\about10\%$).  This is in
accord with observations of other clusters with minimal substructure.
Given the X-ray temperature of A262, one would estimate from the model
results presented in BMM95 that the \bspec of A262 would be $\about0.7$.
The error on this quantity could be substantial, so it is difficult to
ascertain whether or not there is a statistically significant
disagreement with our results.  It is possible that the flat entropy
profile of A262 is commensurate with very mild overheating of gas with
respect to galaxies, rather than the gross overheating argued for by
DJF96.

The fact that galaxy-gas feedback models predict the entropy profile, but
may be inconsistent with typical \bspec values near unity -- indicative of
either non-existent or mild overheating of the gas -- is a conundrum.
This conundrum prompted us to look into the entropy-floor
model in more detail, independent of the A262 study.  We have found
significant new support for the model in that it provides a theoretical
basis for understanding a recently discovered empirical relationship
between $r_{cx}$, $r_v$, and \bgas for a given cluster (NA99).
In NA99 a quadratic correlation was discovered between \bgas and
$r_{cx}/r_v$.  A two parameter empirical
fit was utilized which involved an overall normalization and a cluster
scale length.  As explained below we have produced a fit to the correlation
using the data set of NA99, obtaining it at higher statistical
significance and without removing any select clusters (as was done
in NA99). Moreover the entropy floor model requires only one, not two free
parameters.

\begin{figure}[h]
\plotone{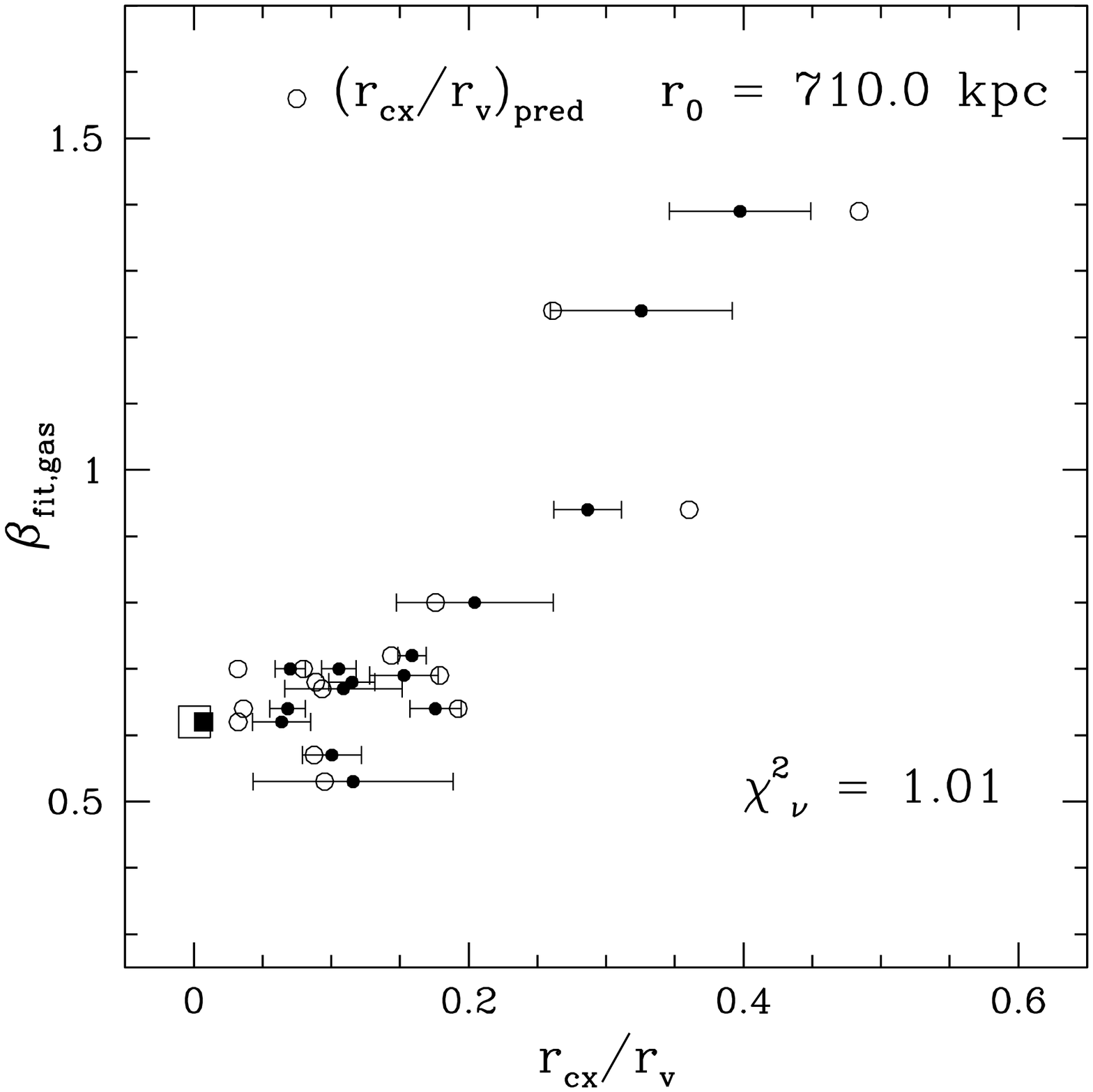}
\figcaption[beta_rcrv.eps]{
$r_{cx}/r_v$ vs \bgas from NA99 (filled circles) and predicted
from the entropy floor model with one free parameter (open circles).  
The error-bars plotted are unscaled.  The observed value for A262 is
indicated with a filled square and its predicted value with the open
square.
\label{beta_rcrv}}
\end{figure}

To obtain this correlation we used the entropy-floor model relation as
given in \citet{evr91} for the central gas density scaling with
temperature, $\rhogas(0)\;\prop\;T^{1/(\gamma - 1)}$, along with an
expression for the virial density at radius $r_v$,
\begin{equation}
	\bar{\rho_v}(z,\omega0) = \frac{3}{r_v^3} \int_0^{r_v}f_b^{-1}
	\frac{\rhogas(0)}{[1+(r/r_{cx})^2]^{3\bgas/2}}\;r^2 dr,
\label{eq7a}
\end{equation}
where $r_{cx}$ is the core radius and $f_b$ the baryonic mass fraction.  This
relation can be manipulated to obtain a relation
between $r_{cx}$, $r_v$, and \bgas:
\begin{equation}
	\left(\frac{r_{cx}}{r_0}\right)^3 \int_0^{r_v/r_{cx}} 
	\frac{y^2}{(1 + y^2)^{3\bgas/2}}\;dy = 1.
\label{eq7}
\end{equation}
In this equation, $r_0$ is a scale length which depends on the
thermodynamics of the energy input and cosmological model (largely
irrelevant for the NA99 sample which was at low redshift).  Equation
(\ref{eq7}) provides a definite relation between the quantities of relevance,
and the only fit parameter, the scale length. In addition, NA99 were
unable to definitively conclude whether the correlation they deduced was
simply between \bgas and $r_{cx}$ or $r_{cx}/r_v$.  The entropy-floor model
clearly indicates that the relationship is between \bgas and the
radius ratio.  We have compared the predictions of the entropy-floor
model in this case directly with the cluster data set of NA99.  We have
followed exactly their procedure for assessing the significance of the
correlation.  In particular, we used $\chi^2$ fitting to
assess the significance of the results and we increased the error bars
on the observational quantities in the manner prescribed by NA99 to
account for the highly correlated errors between $r_{cx}$ and \bgas.
Figure \ref{beta_rcrv} presents the data overlaid with the model predictions.  
Our best fits indicate that the model and data agree at better than the 
99\% confidence level.  This agreement is much better than
the $\about65\%$ confidence level obtained in NA99. Moreover, 
we used all the cluster
data and one less free parameter in our fit.  Even the $\about65\%$
confidence level of NA99 was only obtained when 4 clusters were dropped
from the data set.  Our relation is more complicated than just a
quadratic relation, but it also rests on a firmer theoretical foundation.
The scale length $r_0$ that we derive is, within fitting uncertainties,
comparable to that of the 2 parameter fit of NA99.

The NA99 sample consists of fairly high temperature ($> 3$ KeV) clusters.
Thus, the assumption of the entropy-floor model that the luminosity is
dominated by bremsstrahlung and not line emission is certainly valid.
However, the sample of NA99 does not constitute a homogeneous class of
clusters and one might expect more detailed fitting, perhaps with
different scale lengths, to be required. NA99 pointed out the need for
a cluster sample covering a wider range of temperature and richness
class.  Viewed in this fashion, the above derived relation within the
entropy-floor model seems very encouraging and surprisingly robust.
NA99 argue that the regularity they find in gas density profiles outside
the cluster cores, independent of temperature, is evidence that 
non-gravitational heating is negligible in their sample.  They point out
that this is consistent with the recent work of \citet{pon99},
who demonstrated that lower T clusters have higher
entropies.  However the entropy-floor model assumes a non-gravitational
heating effect that decreases with increasing temperature. The fact
that equation (\ref{eq7}) fits the relation between \bgas, $r_{cx}$ and 
$r_v$ fairly
well, regardless of temperature, validates the idea of substantial
early heating of the proto-intracluster medium (as do flat entropy
profiles in low richness class clusters). The cluster sample does not
go to low enough temperatures to determine whether or not the derived
relation will also hold for lower temperature, lower richness class
clusters. For these clusters the heating is relatively stronger but
this effect is perhaps offset by the increased dominance of line
emission cooling compared to bremsstrahlung. Nevertheless we show the
A262 data point in Figure \ref{beta_rcrv}, and the prediction of equation 
(\ref{eq7}).  The agreement is good.

\section{CONCLUSIONS} \label{conclusions}

1) The early-- and late-type galaxy populations in A262 both have isothermal
VDPs.  There is no evidence that the early-type galaxies have a much lower 
velocity dispersion than the late-type galaxies.  The velocity dispersion 
of the early populations is 569 \kms, much higher than the 330 \kms\  used in 
a previous solution to the $\beta$--problem, and based on a sample of 
only 9 elliptical galaxies.  The main population of late-type galaxies do 
differ from the early-type galaxies in having a non-Gaussian velocity 
dispersion, a different heliocentric velocity and a spatially extended profile.

2) The \bspec derived for A262 on the basis of improved velocity dispersion
and X-ray gas temperature measurements is $1.13\pm0.12$,
consistent with the global mean for clusters with minimal
substructure.  The gas temperature is not a factor of 2 higher than the
galaxy temperature, as previous measurements had suggested, but rather
comparable.  Therefore there is no evidence for strong galaxy-gas
feedback.  The new observations may be in better accord with models
since they do not require artificially high wind velocities or strong
dynamical friction to produce a dramatic factor of 2 difference in the
gas and galaxy temperature.

3) To within the observational uncertainties there is no $\beta$--problem
in A262.  For a velocity anisotropy parameter of $A=0$ the galaxy
temperature is slightly higher than the gas temperature ($\about10\%$),
although within the observational uncertainties the galaxy temperature
could be slightly less than the gas temperature by an almost
comparable amount ($\about5\%$).  Even for reasonable anisotropies 
the best fit gas and galaxy
temperature are also approximately the same.  The important point is
that the $\beta$--problem is solved in a fashion which does not require
significant gas heating with respect to galaxies.  The derived \bspec
is in much better agreement with that obtained from a statistical analysis of
\bspec for either X-ray selected clusters or optically selected
clusters with substructure corrections applied.

4) We have discovered a population of unclassified, low luminosity
galaxies with an isothermal VDP but with a much lower velocity
dispersion than the standard cluster galaxy populations.  These
galaxies have a roughly uniform spatial distribution unlike
the main virialized early-type galaxy population, which is centrally
concentrated.   They may represent a second population of infalling
late-type galaxies or a `local field' population which distinguish themselves 
from the primary
population only by their smaller velocity dispersion.  They are probably 
associated with the Pisces-Perseus ridge in which A262 is located.

5) We have used the entropy-floor model to develop a relation that
explains the recently discovered correlation between cluster \bgas
and the ratio of $r_x$ to $r_v$.  The correlation can be
explained with a one parameter fit, rather than the two parameter
empirical fits previously employed.  This provides additional support
for the entropy-floor model in addition to its ability to approximately
predict the $L_x-T$ relation and the fact that flat entropy profiles are
found in low richness class clusters.

\acknowledgements

We acknowledge useful discussions with Jacqueline vanGorkom and David J.
Helfand.  We also acknowledge the useful comments from the anonymous
referee.  Optical observations were made using the Automated Multi-Object
Spectrometer at the University of California's
Lick Observatory.  This research has made use of data obtained from the 
High Energy
Astrophysics Science Archive Research Center (HEASARC), provided by NASA's
Goddard Space Flight Center.  We have also used NASA's Extra-galactic
Database, ADS Abstract Service, and the Digitized Sky Survey at STScI.

%
%
\begin{deluxetable}{rlrrrrrrl}
\tablecaption{New Velocities for A262\tablenotemark{a}.
	\label{tabobs}}
\tablewidth{0pt}
\tablehead{
\colhead{N}	& \colhead{ID}	& \colhead{RA(J2000)} & \colhead{Dec(J2000)} &
\colhead{cz} 	& \colhead{R} 	& \colhead{N lines}   & \colhead{Type} &
\colhead{Ref} \\
		& & \colhead{h m s} & \colhead{$^\circ$ ' ''} & \colhead{\kms}& & & & 
}
\renewcommand{\arraystretch}{0.96}
\startdata
  1 & -                &  01 50 33.40 &+36 16 42.5 & $  4912 \pm  64 $ &  3.5 &  - &   -2 & \\
  2 & -                &  01 50 45.83 &+36 12 31.5 & $  9750 \pm  39 $ &  3.6 &  - &   -7 & \\
  3 & IC 1732          &  01 50 47.89 &+35 55 57.7 & $  4798 \pm  16 $ & 15.4 &  - &   -2 & CFA\\
  4 & -                &  01 50 51.78 &+35 40 33.6 & $  4817 \pm  37 $ &  3.4 &  - &   -2 & \\
  5 & -                &  01 50 56.76 &+36 08 41.0 & $  3280 \pm  51 $ &  2.7 &  - &   -2 & \\
  6 & -                &  01 51 26.56 &+35 56 03.1 & $  7799 \pm  17 $ &  7.9 &  - &   -7 & \\
  7 & 01485+3549       &  01 51 29.20 &+36 03 57.5 & $  5311 \pm  05 $ &  5.8 &  6 &    5 & CFA\\
  8 & A0148+3537       &  01 51 32.48 &+35 52 32.2 & $  4706 \pm  25 $ &  4.3 &  - &   -7 & CFA\\
  9 & -                &  01 51 51.70 &+36 15 01.8 & $  3755 \pm  33 $ &  5.4 &  - &   -7 & \\
 10 & A0148+3614       &  01 51 53.26 &+36 29 08.3 & $  4297 \pm  18 $ &  3.7 &  - &   20 & CFA\\
 11 & NGC 700          &  01 52 12.74 &+36 05 50.6 & $  4255 \pm  15 $ &  6.0 &  - &   -2 & CFA,SGH98\\
 12 & 01493+3547       &  01 52 16.86 &+36 02 12.2 & $  4590 \pm  13 $ &  9.0 &  - &   -2 & CFA,SGH98\\
 13 & UGC 1339         &  01 52 24.85 &+35 51 23.0 & $  4059 \pm  10 $ & 17.2 &  - &   -2 & CFA\\
 14 & A0149+3615       &  01 52 28.03 &+36 29 52.3 & $  4407 \pm  29 $ &  4.3 &  - &   20 & CFA\\
 15 & -                &  01 52 34.14 &+36 03 09.5 & $  2820 \pm  31 $ &  3.9 &  - &   -7 & \\
 16 & 01497+3615       &  01 52 34.73 &+36 30 03.0 & $  4170 \pm  09 $ &  4.7 &  - &    1 & CFA\\
 17 & NGC 704A         &  01 52 37.71 &+36 07 36.6 & $  4728 \pm  06 $ &  6.8 &  - &   -2 & CFA\\
 18 & NGC 709          &  01 52 50.66 &+36 13 24.3 & $  3781 \pm  37 $ &  8.0 &  - &   -2 & CFA\\
 19 & NGC 710          &  01 52 53.94 &+36 03 11.5 & $  6132 \pm  07 $ &  4.7 &  5 &    5 & CFA\\
 20 & 01500+3615       &  01 52 57.46 &+36 30 46.2 & $  5020 \pm  09 $ &  4.9 &  - &    3 & CFA\\
 21 & A0150+3551B      &  01 52 59.59 &+36 06 26.1 & $  4025 \pm  24 $ &  6.4 &  - &   -7 & CFA\\
 22 & -                &  01 53 04.07 &+35 51 01.7 & $ 10876 \pm  05 $ &  4.2 &  5 &   20 & \\
 23 & -                &  01 53 04.92 &+36 35 24.4 & $  5462 \pm  27 $ &  5.0 &  - &   -2 & \\
 24 & A0150+3606A      &  01 53 09.47 &+36 20 44.6 & $  5045 \pm  23 $ &  4.8 &  - &   20 & CFA\\
 25 & 01504+3546       &  01 53 23.82 &+36 00 43.8 & $  4820 \pm  18 $ &  6.4 &  - &   -5 & CFA,SGH98\\
 26 & NGC 714          &  01 53 29.66 &+36 13 16.6 & $  4418 \pm  16 $ &  5.4 &  - &    0 & CFA,SGH98\\
 27 & 01509+3606       &  01 53 50.15 &+36 21 01.1 & $  4156 \pm  06 $ &  3.0 &  3 &   -5 & CFA,SGH98\\
 28 & -                &  01 53 51.49 &+36 12 11.9 & $  4295 \pm  36 $ &  3.7 &  - &   -7 & \\
 29 & 01513+3540       &  01 54 11.32 &+35 55 20.5 & $  4285 \pm  18 $ &  4.9 &  5 &    1 & CFA\\
 30 & -                &  01 54 47.50 &+36 01 20.4 & $  4259 \pm 104 $ &  4.4 &  - &   -7 & \\
 31 & -                &  01 54 51.54 &+35 57 32.4 & $  2859 \pm  23 $ &  4.9 &  - &   -2 & \\
 32 & -                &  01 54 54.28 &+36 01 57.8 & $  3545 \pm  19 $ &  6.7 &  - &   -2 & \\
 33 & A0151+3537       &  01 54 55.05 &+35 52 28.2 & $  4886 \pm  47 $ &  4.1 &  4 &   20 & CFA\\
\enddata
\tablenotetext{a}{
Column 1 gives our sequential ID, column 2 gives the published ID, if
available, column 3 and 4 give the coordinates, column 5 gives the heliocentric
velocity with 1$\sigma$ error, column 6 gives the \citet{ton79} R
value for the absorption-line galaxies, column 7 gives the number of
emission lines fit for emission-line galaxies, column 8 gives the
morphological type code \citep{huc99}, and column 9 gives the
reference for previously published observations.
}
\end{deluxetable}

%
%
\begin{deluxetable}{lrrrrrrr}
\tablecaption{Derived Optical Properties out to 
	$r_v$ ($1\degpnt17$)\tablenotemark{a}. 
	\label{tabvel}}
\tablewidth{0pt}
\tablehead{
	\colhead{Pop} & \colhead{N$_{gal}$} & 
	\colhead{cz} & \colhead{\sigp} & \colhead{TI} & 
	\colhead{a-test} & \colhead{W-test} \\
	& & \colhead{\kms} & \colhead{\kms}
}
\startdata
All	& 101	& $4869\pm55$	& $548\pm36$	& 0.967	& 0.804	& 0.983 (66\%)\\
Early	& 38	& $4736\pm94$	& $569\pm61$	& 0.973	& 0.799	& 0.985 (92\%)\\
Late	& 40	& $4993\pm92$	& $572\pm50$	& 0.861 & 0.832 & 0.962 (28\%)\\
Unclass	& 23	& $4894\pm90$	& $418\pm74$	& 1.036 & 0.766 & 0.978 (85\%)\\
\enddata
\tablenotetext{a}{
Column 1 gives the population type, column 2 gives the number of galaxies
in the sample, column 3 gives the average heliocentric velocity with
$1\sigma$ error, column 4 gives the velocity dispersion with $1\sigma$
error, column 5 gives the scaled tail index, column 6 gives the a-test
statistic, column 7 gives the W-test statistic with percent probability in
parenthesis.
}
\end{deluxetable}

%
%
\begin{deluxetable}{lllll}
\tablecaption{Comparison of $\beta$s.
	\label{tabbetas}}
\tablewidth{0pt}
\tablehead{
	\colhead{Source} & \colhead{\bspec} & \colhead{\bfit} & 
	\colhead{\bgas} & \colhead{\bgal} \\
}
\startdata
\multicolumn{5}{c}{A262}\\
DJF96	 & 0.51	         & 0.53	        & 0.53	      & 1.0	\\
This work& $1.13\pm0.12$ & $0.82\pm0.1$	& $0.6\pm0.1$ & 0.73\tablenotemark{a}\\
\hline
\multicolumn{2}{c}{Statistical \bspec} & & &\\
BMM95\tablenotemark{b}	 & $1.14\pm0.08$  & & & \\
ES91\tablenotemark{c}	 & $0.91^{+0.11}_{-0.13}$  & & & \\
\enddata
\tablenotetext{a}{from G98}
\tablenotetext{b}{calculated from Table 1 in \citet{lub93}}
\tablenotetext{c}{\citet{edg91b}}
\end{deluxetable}

\clearpage

\end{document}